\documentclass[11pt]{article}
\usepackage{graphicx}
\DeclareGraphicsExtensions{.eps,.ps}
\usepackage{epsfig, graphicx}

\usepackage{subfigure}
\newcommand{\be}[1]{\begin{equation} \label{eq#1} }
\newcommand{\ee}{\end{equation}}
\newcommand{\bea}[1]{\begin{eqnarray} \label{eq#1} }
\newcommand{\eea}{\end{eqnarray}}

\def\setb@se#1{\baselineskip=#1 \normalbaselineskip=#1}
\def\captionfont{\setb@se{10pt}\protect\footnotesize}
\begin{document}
\mbox{}

\vskip0.0cm
\begin{center}

{\large Analysis on Irreversible Processes using the Phase-Field Variational Approach with the Entropy or Energy Functional \\}
\vskip0.7cm  Peng Zhou$^1$\\[0.4cm]
$^1$Department of Astronautics Science and Mechanics, Harbin Institute of Technology, Harbin, \\
    Heilongjiang, 150001, P.R. China  \\
    Email: zhoup@hit.edu.cn
\vskip0.0cm

\end{center}
\vskip0.7cm

\begin{abstract}
The variational approach usually used in phase field models (PFVA) is applied here to analyse complex irreversible processes such as thermoelectric (TE) effects and thermally driven mass transport (TDMT). Complex irreversible processes arise from the coupling effects between simple irreversible processes. Each simple irreversible process is described by the evolution of a field variable and there is an entropy or energy density function associated with it. During complex irreversible processes with multiple fields present, this entropy or energy density function is assumed to be dependent on all independent field variables. Using the total entropy functionals, the TE effects and TDMT are analysed and important kinetic coefficients such as the Seebeck coefficient and the heat of transport are determined with straightforward physical contents. Using the total energy functionals, the linear irreversible processes are analysed with the Onsager approach and the nonlinear irreversible processes with PFVA. It is found both the Onsager's relations and the fluxes defined using PFVA guarantee the satisfaction of the first law of thermodynamics during the process of conversion of energies. In order to analyze the diffusion process under the influence of elasticity, PFVA is also modified to incorporate the reversible evolution of elastic fields. It is shown that energies are conserved via both the irreversible diffusion process and the reversible evolution of the elastic fields. In the end, PFVA is generalized to study nonequilibrium thermodynamics using an extra kinetic contribution to the entropy density function. The analyses can be extended to a nonequilibrium thermodynamic system with multiple physical fields present. Thus, it is believed PFVA has the potential of not only significantly advancing our understanding of the thermodynamics of irreversible processes, but also making thermodynamics as a discipline and the study of it truly dynamic.

\end{abstract}

\vspace{0.7cm}


\vspace{0.7cm}

\noindent \textbf{Update: Section 6 is a new section which focuses on the application of PFVA in nonequilibrium thermodynamics.} 

\newpage

\section {Introduction}

Thermodynamic processes are classified as reversible and irreversible processes, however, all natural processes are in fact irreversible. Typical and simple irreversible processes, such as heat transfer, friction and mixing processes in multi-component solutions, have been studied extensively and thus become well-known. But for more complex irreversible processes, e.g., thermomigration, thermoelectric effects and electromigration, their mechanisms are still not thoroughly understood and thus arouse interests of many researchers. These complex irreversible processes are actually coupling effects between two simple irreversible processes. For examples, thermomigration (electromigration) is the atomic diffusion due to the presence of a temperature gradient (electric currents) and thus is the coupling effect between atomic diffusion and thermal (electrical) conduction. Thermoelectric effects are coupling effects between thermal conduction and electrical conduction. In general, irreversible processes can involve three or even more simple irreversible processes, which sometimes include chemical reactions. These irreversible processes are not only important to industrial applications, but also vital to the survival of living organisms. Thus, in the beginning it is helpful to review the definition of reversible and irreversible processes in thermodynamics. Reversible processes are in fact idealized processes which only happen at an infinitesimal rate of change. Any thermodynamic system, underwent certain changes and was restored to its
initial configuration both via reversible processes, results in no entropy change for both the system and the surroundings and thus, no permanent changes in the
universe. However, all natural processes happen at a finite rate of change and thus are irreversible processes. Irreversible processes are accompanied
by both the dissipation of free energy and the production of entropy. Though the thermodynamic system can be restored to its initial state via irreversible processes, the entropy of the surroundings is increased which indicates the universe itself is already different.$^{\cite{DeHoff:1993}}$ An alternative and more specific definition of reversible and irreversible processes
is related to the time reversal transformation of their governing equations. Governing equations which are invariant
under the time reversal transformation are associated with reversible processes, otherwise with irreversible processes.$^{\cite{Prigogine:1967}}$

The study of irreversible processes dates back to centuries ago and two major approaches were developed to study them. The first
approach is directly based on the well-known Boltzmann equation,$^{\cite{Ziman:1999}}$ which is an integro-differential equation of the distribution
 function $f$ of particles. The function $f$ first needs to be linearized near the equilibrium state to introduce the driving forces
 (e.g., thermal, electrical and so on) into the equation. In order to obtain an analytical and linearized expression of $f$, further simplification and assumption needs to be made about the rate of change of $f$ due to the scattering effects. Then, the thermal and electrical fluxes are determined using the linearized expression of $f$, and the coupling coefficients are found to be related to an integral over the Fermi surface. The second resorts to the
 framework developed by Onsager.$^{\cite{Onsager:1932}}$ This approach assumes that the thermodynamic system is quite close to equilibrium, so that the
 fluxes are defined to be linearly dependent on all driving forces involved. Using the principle of microscopic reversibility, the coupling coefficients
 are shown to be related as $L_{ij}=L_{ij}$, i.e., the so-called Onsager's reciprocal relations. The rate of production of entropy is found to be
 the summation of the products of the fluxes with their corresponding driving forces. The factor $\frac{1}{T}$ is usually included in the driving forces.
 However, deficiencies of both approaches are evident.
 For the Boltzman approach, several bold assumptions are involved to obtain the linearized expression of $f$, and the coupling coefficients given by it
 usually contain the energy of the Fermi surface, which makes interpretations of the physical contents of these coefficients obscure. For the Onsager
 approach, the coupling coefficients are assumed constants and it is certainly not helpful for their physical interpretations; moreover, the reciprocal
 relations are often questioned owing to the lack of a sound proof at the macroscopical level. Furthermore, both approaches are only applicable to
 irreversible processes with small deviations from equilibrium so that the use of linearizations is valid.

In this paper, the variational approach, which is commonly used in phase field models to construct governing equations, is applied to analyze irreversible
processes. For brevity, this approach is here called the phase field variational approach (PFVA). Phase field models have been developed to study many types of problems with important applications.$^{\cite{Chen:2002}}$ Usually, these models use the Landau-Ginzburg free energy functional to construct governing equations via a variational method, which guarantees the total
free energy of an isolated system decreases along solution paths. The free energy functional consists of bulk free energies and gradient energies.$^{\cite{Cahn:1961}}$ The former
includes chemical and elastic free energies and so on. The latter represents interfacial and surface energies. Formulation of these models uses the iso-thermal
 condition and thus the internal energy of the system can be treated as constant. However, in certain important industrial problems such as solidification
 and thermomigration, the temperature field is no longer uniform and usually certain temperature gradients are maintained. In order to model these problems,
 improvements were made to use the entropy functional to construct governing equations.$^{\cite{Penrose:1990}}$ The fluxes are defined to be proportional to the gradients of the variational derivatives of the entropy functional. This guarantees the local rate of the entropy production to be nonnegative and thus
 the second law of thermodynamics is satisfied. The improved formulations were soon applied in phase field models for solidification problems.$^{\cite{Wang:1993,Wheeler:1996}}$ For the irreversible processes studied here, it will be shown with PFVA the use of linearizations in derivations is no longer needed, and also expressions of the coupling coefficients contain straightforward physical interpretations.

In irreversible processes, thermoelectric (TE) effects and the phenomenon of the thermally driven mass transport (TDMT) are two important examples with
many applications. Also diffusion under the influence of elastic fields is another process heavily studied. The majority of this paper is devoted to the analysis of these processes. The purpose is three-fold: first, to determine the kinetic coefficients for the coupling effects and have their physical contents explicitly shown and thus, contribute to a straightforward understanding of both TE effects and TDMT;
second, using macroscopic formulations to reveal the underlying physical foundations of Onsager's reciprocal relations and thus to
explain the conversion of energies in both irreversible processes; finally, using these examples to show that PFVA can be developed into a general
approach to analyse coupling effects in irreversible thermodynamics. Organization of this paper is as follows. In the second and third sections, TE effects and TDMT are analysed using PFVA with the entropy functionals, respectively. In the fourth section, the dissipation and conversion of free energies are discussed using PFVA with the free energy functional. Diffusion under the influence of elastic fields is analysed in the fifth section. And a summary is presented in the end.

\section{Thermoelectric Effects}

Thermoelectric effects arouse the interests of many researchers owing to their promising
potentials in the area of energy conversion.$^{\cite{Tritt:2011}}$ Endeavors are devoted to improve the
intrinsic property of TE materials$^{\cite{DiSalvo:1999}}$ and to broaden the applications of TE devices as well.$^{\cite{Bell:2008}}$ Thermoelectric effects, consisting of the Seebeck effect, Peltier effect and Thomson effect, refer to the coupling influences
between the thermal and electrical conductions. The Seebeck effect is the
cause of an electrical current due to the existence of a temperature gradient in the conductive loop. Thus, the electric
current density $\vec{J}$ with the seebeck effect present is $\vec{J}=-\sigma (\nabla \phi + S_e \nabla T)$, where $\sigma$ is the
conductivity, $\phi$ is the electric potential, $T$ is temperature and $S_e$ is the Seebeck
coefficient or the thermal power. When an electric current flow through the junction of two conductors A and B, heat is either
released or absorbed locally and this is the Peltier effect. The rate of change of heat $Q$ is given by $Q_t =(\Pi_A-\Pi_B)I$
where $\Pi_i$ $£¨i=A, B£©$ are the Peltier coefficients of the corresponding conductors and $I$ is the electric current.
The Thomson effect gives more realistic description of the Peltier effect since the material coefficients are not
only allowed to vary at the junction but also throughout the bulk of the material. Thus, the rate of change of the heat density $q$ is
given by $q_t =-\kappa \vec{J}\cdot \nabla T$, where $\kappa$ is the Thomson coefficient. Analysis showed that these material
coefficients abide by the first and the second Thomson's relations, i.e., $\Pi=S_e T$ and $\kappa = \frac{d\Pi}{dT}-S_e$,
respectively. Onsager's relations contributed to the proof of the second relation. Traditionally, these effects
were analysed via both the Boltzman approach and the Onsager approach. Interested readers are directed
to these references for a pedagogical treatment$^{\cite{Mott:1979}}$$^{\cite{Dommelen:2014}}$, a more advanced treatment using the linearized Boltzman's equation,$^{\cite{Ziman:1999}}$ and a detailed review for related theoretical work.$^{\cite{Bulusu:2008}}$ Conclusions based on the Mott formula$^{\cite{Tritt:2011}}$$^{\cite{Mott:1979}}$ are quoted here as references.
The seebeck coefficient $S_e$ was found to be $-\frac{k_B}{\sigma e} \int \frac{E-\mu}{k_B T} \sigma(E) [-\frac{df(E)}{dE}] dE$, where $k_B$ is
the Boltzman constant, $E$ is the energy of electrons, $\mu$ is the energy of the Fermi surface and $f(E)$ is the
Fermi-Dirac distribution function. In metals, $S_e$ can be rewritten as $-\frac{\pi^2 k_B^2 T}{3 e} \frac{\sigma'(\mu)}{\sigma}$ and finally simplified to be
$- \frac{\pi^2 k_B}{3 e} \frac{T}{T_F}$ if the electrons are treated as Fermi gas, where $T_F$ is the Fermi temperature. In semiconductors,
$S_e=\frac{\sigma_c S_{ec}+\sigma_v S_{ev}}{\sigma_c+\sigma_v}$ where $S_{ec}= - \frac{k_B}{e} \frac{ E_{c} - \mu}{k_B T} + a_{c} + 1$ ($n$-type), $S_{ev}= \frac{k_B}{e} \frac{- E_{v} + \mu}{k_B T} + a_{v} + 1$ ($p$-type), and $a_{c,v}$ are constants between 1 and 3.
The efficiency of a TE material to produce power is measured by its figure of merit, $\frac{\sigma S_e^2 T}{K}$, where $K$ is the
heat conductivity. This dimensionless quantity guided the search for better TE materials and nowadays many progresses have been made.$^{\cite{Sootsman:2009}}$
In the following two subsections, PFVA with the entropy functional is applied to analyse both the Seebeck and Peltier effects. For simplicity, analyses are performed in a single phase so that contributions from the gradient entropy terms can be ignored. However, this has no influences on the important
kinetic coefficients to be determined.

\subsection {The Seebeck effect}

Consider a thermodynamic system, either a metal or a semiconductor, where thermal
and electrical conductions coexist. For the thermal conduction, the thermal entropy per volume was
defined to be $S_q=\int_0^T \frac{C_P(\theta)}{\theta} d \theta$ where $C_P$ is the vomlumetric heat
capacity at constant pressure. Two major heat carriers contributed to the heat capacity in metals and
semiconductors are phonons and electrons or holes. Here, discussion is performed in metals or the n-type semiconductors
so that electrons are the conducting species electrically. Let $n$ be the density of electrons, then $C_P$ should also be a function
of $n$. Thus rewrite the thermal entropy as  $S_q=\int_0^T \frac{C_P(\theta, n)}{\theta} d \theta$. The entropy
function related to the electric conduction, thought not well studied, can be found as
$S_n =-\frac{\partial [n (-e) \phi]}{\partial T}$ using the relation $S=-\frac{\partial G}{\partial T}|_{P}$,
where $G$ is the Gibbs free energy density of the system. As a result, the total entropy density function $S$ of this system is the summation of
the above two entropy functions,
\begin{equation}
S(T,n) = \int^T_{0} \frac{C_P(\theta,n)}{\theta} d \theta + \frac{\partial (n e \phi)}{\partial T}. \label{eq:m1}
\end{equation}

Thermal and electric conduction are both irreversible processes and they must abide by the generalized second law of thermodynamics, i.e.,
the local rate of production of entropy is everywhere non-negative. Furthermore, the gradients of temperature and electrical potential in the whole system are assumed moderate thus the assumption of local thermodynamic equilibrium is valid. Conduction processes are first considered in metals with the electron density
 $n$ being constant. Then, analyses are extended to semiconductors in which $n$ is temperature dependent. Note that the definition of the electrical entropy function $S_n$ is a very preliminary one. For simplicity, analysis on it is performed in one-dimensional case.
 Both the electrical potential $\phi$ and temperature $T$ are functions of position $x$, thus $\phi=\phi(x)=\phi[x(T)]=\phi(T)$. Hence, the second term in Eqn (\ref{eq:m1}), $S_n=\frac{\partial (n e \phi)}{\partial T}=ne\frac{\partial \phi}{\partial x} \frac{\partial x}{\partial T}=-neRJ\frac{\partial x}{\partial T}$, where $R$ is the resistivity. Here Ohm's law, $J=\sigma E$, is used and $J$ only refers to the portion of the current density due to the electrical field. (The other portion arises from the temperature gradient.) When the system achieves a steady state, then both $\frac{\partial \phi}{\partial x}$ and $\frac{\partial T}{\partial x}$ are constants, indicating $J\frac{\partial x}{\partial T}$ is constant. The temperature-dependence of $S_n$ is further used in the derivation below. Here it is assumed this dependence mainly arises from the resistivity $R$ and the dependence of the product $J\frac{\partial x}{\partial T}$ is minor and can be neglected. As a result, the corresponding entropy functional is
\begin{equation}
\textbf{S} = \int_V S(T,n) dv = \int_V [ \int^T_{0} \frac{C_P(\theta,n)}{\theta} d \theta -neRJ\frac{\partial x}{\partial T} ] dv,
\end{equation}
and its rate of change is
\begin{eqnarray} \label{eq:1}
\frac{d\textbf{S}}{dt} &=& \int_V \frac{\partial S}{\partial t} dv \nonumber \\
 &=&  \int_V \left \{ [\int^T_0 \frac{\partial C_P(\theta,n) / \partial n}{\theta} d \theta -eRJ\frac{\partial x}{\partial T}] n_t  + [\frac{C_P(T,n)}{T} -ne J\frac{\partial x}{\partial T} \frac{\partial R}{\partial T}  ] T_t \right \} dv \nonumber \\
 &=&  \int_V \left\{ \Lambda n_t + \Omega C_P T_t  \right \} dv,
\end{eqnarray}
where $\Lambda$ and $\Omega$ are introduced for brevity. The equations of conservation of electrons and heat $Q$ are
\begin{eqnarray}
\frac{\partial n}{ \partial t} &=&  - \nabla \cdot \vec{J_n}; \label{eq:2} \\
\frac{\partial Q}{ \partial t} &=& C_P \frac{\partial T}{ \partial t} = - \nabla \cdot \vec{J_q}. \label{eq:3}
\end{eqnarray}
Substitute Eqns (\ref{eq:2}) and (\ref{eq:3}) into Eqn (\ref{eq:1}) and using integration by parts, we have
\begin{eqnarray}
\frac{d\textbf{S}}{dt} &=& \int_V \left\{ \vec{J_n} \cdot \nabla \Lambda + \vec{J_q} \cdot \nabla \Omega \right\} dv \nonumber \\
  & &- \oint_A \left\{ \Lambda \vec{J_c}  + \Omega \frac{\vec{J_q}}{T} \right\} \cdot \vec{n} da.
\end{eqnarray}
In the above formula, the first integrand in the volume integral is the local rate of
the entropy product and the second integrand in the surface integral is the local rate of the entropy exchanged with the surroundings.
Thus, to guarantee the local rate of entropy production to be positive, the fluxes can be assumed to be
\begin{eqnarray}
 \vec{J}_n &=& M_n(n,T) \nabla \Lambda \\
 \vec{J}_q &=& M_q(n,T) \nabla \Omega \label{eq:7}
\end{eqnarray}
where $M_n(n,T)$ and $M_q(n,T)$ are proportional coefficients which are related to diffusivity and thermal conductivity. The
heat flux equation will be discussed in the next subsection and here we focus on the electron flux equation first. Expansion and simplification of the
electron flux equation after the substitution of $\Lambda$ give
\begin{equation}
 \vec{J_n} = M_n(n,T) [ \frac{\partial C_P}{\partial n} \frac{1}{T} \nabla T + \int^T_{0} \frac{\partial^2 C_P(\theta,n) / \partial n^2}{\theta} d \theta \nabla n] -M_n(n,T) e J\frac{\partial x}{\partial T} \frac{\partial R}{\partial x} \vec{i_x} \label{eq:4}
\end{equation}
The second term in the square bracket can be dropped assuming the contribution of electrons to $C_p$ is mostly linear so that $\frac{\partial^2 C_P(n, \theta)}{\partial n^2}$ is negligible; furthermore, electrons are highly mobile thus $\nabla n \approx 0$. The last term can
 be rewritten as $-M_n(n,T) e \frac{\partial R}{\partial x} \frac{\partial x}{\partial T}  J  \vec{i}_x= - M_n(n,T) e \frac{\partial R}{\partial T} \sigma \vec{E} $. Given $\nabla T =0$, then $J_n= - \sigma \vec{E} = - M_n(n,T) e \frac{\partial R}{\partial T} \sigma \vec{E}$. Thus,
 $M_n(n,T) = \frac{1}{e \partial R/\partial T}$. Hence, Eqn (\ref{eq:4}) is reduced to be
\begin{equation}
\vec{J_n} = - \sigma (-\frac{R}{e \partial R/\partial T} \frac{\partial C_P}{\partial n} \frac{1}{T} \nabla T  + \vec{E}). \label{eq:10}
\end{equation}
Then, the Seebeck coefficient is found to be
\begin{equation}
 S_e = - \frac{R}{e T \partial R/\partial T} \frac{\partial C_P}{\partial n}. \label{eq:9}
\end{equation}
In metals, the heat capacity of electrons Fermi gas at constant volume is found to be $C_v=\frac{1}{2} N_0 \pi^2 k_B \frac{T}{T_F}$ where
$N_0$ is the total number of electrons. It can be taken as a good approximation as $C_P$ in solids. Thus, $\frac{\partial C_P}{\partial n}\approx\frac{\partial C_V}{\partial N_0}=\frac{1}{2} \pi^2 k_B \frac{T}{T_F}$. Substitute it into the above equation, then $S_e=- \frac{\pi^2 k_B}{2 e} \frac{R}{\partial R/\partial T} \frac{1}{T_F}$. It can be seen that its form agrees qualitatively with the expression given in the introduction using the Mott formula; however, there are extra dependence on the resistivity-related terms and the numerical factor is $\frac{1}{2}$ instead of $\frac{1}{3}$.

Now, consider conduction processes in semiconductors. Analyses are performed in n-type semiconductors. Results are given analogously
in p-type semiconductors. In n-type semiconductors, the number density $n$ of conducting electrons is
\begin{equation}
 n = 2 (\frac{m_e k_B T}{2 \pi \hbar^2})^{3/2} exp( \frac{\mu - E_c}{k_B T}).
\end{equation}
Evidently, $n$ is temperature dependent and its derivative w.r.t temperature is
\begin{equation}
 \frac{\partial n}{\partial T} = n (\frac{3}{2}+\frac{E_c-\mu}{k_B T}) T^{-1}.
\end{equation}
Thus, in semiconductors, the entropy related to electrical conduction is
\begin{equation}
 \frac{\partial (n e \phi)}{\partial T} = (\frac{3}{2}+\frac{E_c-\mu}{k_B T}) T^{-1} n e \phi -neRJ\frac{\partial x}{\partial T}.
\end{equation}
Using the same approach above, one extra term is found in the electron flux and it is
\begin{eqnarray}
\vec{J}^{ex}_n &=& M_n(n,T) \nabla [e \phi (\frac{3}{2}+\frac{E_c-\mu}{k_B T})T^{-1}]  \nonumber \\
  &=& M_n(n,T)[-e(\frac{3}{2}+\frac{E_c-\mu}{k_B T})T^{-1}(-\nabla \phi) + e \phi [-\frac{3}{2T^2}-2\frac{E_c-\mu}{k_BT^3}] \nabla T].
\end{eqnarray}
The second term in the square bracket can be dropped since it contains $T^{-2}$ and $T^{-3}$ and can be assumed negligible.
Using $M_n(n,T) = \frac{1}{e \partial R/\partial T}$, this extra term is written as
\begin{equation}
\vec{J}^{ex}_n = - \sigma  [\frac{R}{T \partial R/\partial T} (\frac{3}{2}+\frac{E_c-\mu}{k_B T})] \vec{E}. \label{eq:5}
\end{equation}
Analogously, in p-type semiconductors, the flux for holes can be found to be
\begin{equation}
\vec{J}_{v} = \sigma (-\frac{R}{e \partial R / \partial T} \frac{\partial C_P}{\partial n} \frac{1}{T}) \nabla T + \sigma \vec{E} +\sigma  [\frac{R}{T \partial R/\partial T} (\frac{3}{2}+\frac{\mu-E_v}{k_B T})] \vec{E}. \label{eq:6}
\end{equation}
Note that the coefficients in the square bracket of Eqn (\ref{eq:5}) and that of the last term of Eqn (\ref{eq:6}) qualitatively agree with
the Seebeck coefficients given in the introduction for n-type and p-type semiconductors, however it is associated with the electrical
field $\vec{E}$ rather than $\nabla T$.

From the above derivations, it can be seen that the temperature gradient has two contributions to the electrical currents.
The first one is proportional to $\frac{\partial C_P}{\partial n} \nabla T$. Treating the electrons as Fermi gas, then $\frac{\partial C_P}{\partial n}$ gives the heat capacity of each electron. Assume the system achieved steady state, then $\nabla T = \frac{T_h-T_l}{X_h-X_l}$. Thus, this first contribution can be considered as being proportional to the difference in the thermal energies carried by electrons at the ends with higher and lower temperatures. That is, the difference in the thermal energies, which is substantially the difference in the kinetic energies of electrons at the two ends, contributed to the generation of electric currents. The second contribution only exists in semiconductors and it is proportional $\frac{1}{T}(\frac{3}{2}+\frac{E_{c/v}-\mu}{k_B T})$, which in fact arises from $\frac{\partial n}{\partial T}$. At the steady state, $\frac{\partial n}{\partial T}=\frac{n_h-n_l}{T_h-T_l}$. Thus, the difference in the number density of electrons or holes at the two ends contributed to the generation of electric currents. At the end with higher temperatures, more electrons or holes are excited to the conduction band. Then they are driven down by the electrical field to the end with lower temperatures. To achieve thermal equilibrium, these electrical carriers will eventually return to the forbidden band and release their thermal energies into the surroundings. Since this contribution arises
from the difference in the number density of electrical carriers, it is reasonable that this term is associated with the electrical field. In brief,
for the Seebeck effect, differences in both the kinetic energies and the number densities of electrical carriers contributed the generation of electrical
currents.

\subsection{The Peltier Effect}

In this section, the Peltier Effect in metals is analysed. In metals ,The heat flux in Eqn (\ref{eq:7}) after
substitution of $\Omega$ is
\begin{eqnarray}
\vec{J}_q &=& M_q(n,T) \nabla [\frac{1}{T} -ne J\frac{\partial x}{\partial T} \frac{1}{C_P(T,n)}\frac{\partial R}{\partial T}] \nonumber \\
  &=& - M_q(n,T) \frac{1}{T^2} \nabla T - ne J\frac{\partial x}{\partial T} \nabla (\frac{1}{C_P(T,n)}\frac{\partial R}{\partial T})
\end{eqnarray}
Similarly $M_q(n,T)$ is identified as $KT^2$ where $K$ is the heat conductivity. The term $\nabla (\frac{1}{C_P(T,n)}\frac{\partial R}{\partial T})$
can be expanded as -$\frac{1}{C_P^2}(\frac{\partial C_P}{\partial T} \nabla T + \frac{\partial C_P}{\partial n} \nabla n)+\frac{1}{C_P}\frac{\partial^2 R}{\partial T^2} \nabla T$. Here, $\nabla n \approx 0$ due to the high mobility of electrons. Thus, the heat flux can be rewritten as
\begin{eqnarray}
\vec{J}_q &=& -K \nabla T - K T^2 ne J\frac{\partial x}{\partial T}(\frac{1}{C_P^2}\frac{\partial C_P}{\partial T} \frac{\partial T}{\partial x} \vec{i}_x
+ \frac{1}{C_P}\frac{\partial^2 R}{\partial T^2} \frac{\partial T}{\partial x} \vec{i}_x )  \nonumber \\
  &=& -K \nabla T - K T^2 ne (\frac{1}{C_P^2}\frac{\partial C_P}{\partial T} + \frac{1}{C_P}\frac{\partial^2 R}{\partial T^2}) \vec{J},
\end{eqnarray}
where the Peltier's coefficient is identified as
\begin{equation}
\Pi= - K T^2 ne (\frac{1}{C_P^2}\frac{\partial C_P}{\partial T} + \frac{1}{C_P}\frac{\partial^2 R}{\partial T^2}). \label{eq:8}
\end{equation}
Note, compared to the Seebeck coefficient, the physical content of $\Pi$ given above is not very straightforward. This difficulty, we argue, was caused by the preliminary form of the electrical entropy $s_n=\frac{\partial (n e \phi)}{\partial T}$, as well as the heat capacity $C_p$ in thermal conduction Eqn (\ref{eq:3}). To obtain a neat expression for $\Pi$ with clear physical contents, $S_n$ needs be studied in details so that an explicit expression of this function can be determined to describe the contribution to entropy from the electron flow in the conduction band of metals . Moreover, it could also be helpful to replace Eqn (\ref{eq:3}) with a conduction equation for thermal phonons. A similar approach needs to be adopted in semiconductors.

\section {Thermally Driven Mass Transport}

In this section, the phenomenon of thermally driven mass transport (TDMT) is analogously analysed using PFVA with the corresponding entropy functional.
TDMT refers to the transport of mass under the influence of a
temperature gradient and it has been observed in all three phases of matter. It is usually named thermophoresis in gas mixtures,
the Ludwig-Soret effect in liquid solutions and thermomigration (thermodiffusion or thermotransport)
in solid phases. However, each term also commonly applies to this phenomenon in all three phases.
Study of this phenomenon can date back to more than one and half centuries and detailed narrations can be found in these literatures.$^{\cite{Zheng:2002}}$$^{\cite{Piazza:2008}}$
$^{\cite{Wiegand:2004}}$$^{\cite{Huntington:1975}}$  To begin with, thermophoresis was first reported in aerosol mixtures
 by the British scientist John Tyndall in 1870 while studying the floating particles in air of London.$^{\cite{Tyndall:1870}}$
 The Ludwig-Soret effect was first reported in 1856 by Carl Ludwig$^{\cite{Ludwig:1856}}$ in concentrated salt solutions,
 and was latter studied in details in 1879 by Charles Soret.$^{\cite{Soret:1879}}$ Thermomigration in solid phases, as well
 as electromigration, became a subject of interests due to reliability concerns in integrated circuits around 1950s.$^{\cite{Huntington:1975}}$
 In all three phases of matter, mass transport was observed at the presence of a temperature gradient. For over 100 years, this thermally driven
 mass transport was shown to have important applications in many fields. For examples, in gas phases,$^{\cite{Zheng:2002}}$ thermophoresis repels
 particles away from
surfaces of heated semiconductor wafers to avoid microcontamination. In liquid phases, the Ludwig-Soret effect also affects the global circulation of sea water,$^{\cite{Wurger:2006}}$ and the concentration distribution in crude oil reservoirs;$^{\cite{Platten:2006}}$ recently, it was also
used to study soft matters such as polymer$^{\cite{Yang:2012}}$ or biomolecular solutions.$^{\cite{Buhr:2006}}$ In the
solid phase, thermomigration was shown to cause reliability concerns in flip chip solder joints$^{\cite{Ye:2003}}$ and enrichments of
fuel elements and fission products at different locations within nuclear fuel pellets and formation of brittle phases at the blade of cutting tools.$^{\cite{Hehenkamp:1976}}$
Thus, TDMT is related to both industrial applications and natural processes and involves both microscopic and
macroscopic systems which spans from nanometers to kilometers.

In different phases, description of TDMT varies. The effect of thermophoresis in aerosol and gas mixtures
is represented by the thermophoretic force $f_T$ and velocity $v_T$ of the suspended particles. They are believed to diffuse at
a constant $v_T$ owing to the force balance between $f_T$ and a frictional force from the surroundings.$^{\cite{Piazza:2008}}$
In binary liquid solutions, the mass flux $J$ is written as $J = - \rho D \nabla c - \rho c(1 - c) D_T \nabla T$,
where $\rho$ is the mass density, $c$ is the concentration, $D$ is the diffusion coefficient, $D_T$ is the thermal diffusion
coefficient and $T$ is the temperature.$^{\cite{Wiegand:2004}}$ The Soret coefficient is defined as
$S_T = \frac{D_T}{D}$. In the solid phase, the flux component arising from thermomigration
is written as $-D \, c(1-c)\, \frac{ Q^*}{kT} \, \frac{ \nabla T}{T}$ where $k$ is Boltzman's constant and $Q^*$ is the heat of transport.$^{\cite{Huntington:1975}}$ Many experimental techniques have been developed to investigate this driven mass
transport owing to its practical importance.$^{\cite{Zheng:2002,Piazza:2008,Wiegand:2004,Huntington:1975,Platten:2006}}$
The techniques used to study thermomigration in the solid phase are similar to those for electromigration.$^{\cite{Huntington:1975}}$
Besides experimental techniques, recently numerical simulations especially molecular dynamics simulations have also emerged as
an important tool.$^{\cite{Wiegand:2004}}$

However, theoretically, it is a pity that even nowadays understanding of TDMT is still not very clear.
Compared to those in solid phases,$^{\cite{Philibert:1991}}$ theoretical interpretations in fluid phases are more advanced.$^{\cite{Zheng:2002,Piazza:2008,Wiegand:2004,Talbot:1980}}$
Both the Boltzmann approach and the Onsager approach are used to analyze TDMT. The Boltzmann approach is usually
preferred in fluid phases since microscopically the movement of gas particles or solute species in a non-uniform temperature field is a problem of
non-equilibrium statistical mechanics. However, the Onsager approach is usually preferred in solid phases. Because macroscopically, migration
of atoms or molecules driven by a temperature gradient belongs to the category of kinetics and irreversible thermodynamics,
where analysis usually benefits from diffusion theories and Onsager's formulation. Note that, though the Onsager approach can be universally
applied to all three phases, this approach was sometimes questioned by researchers working on fluid phases.$^{\cite{Piazza:2008,Talbot:1980}}$

In the following subsections, the mechanism of thermally driven mass transport is discussed. As we know, commonly studied mixtures consists of
chemical mixtures and mechanical mixtures. In chemical mixtures, particles refer to interacting atoms and molecules. Usually atoms are bonded
and molecules interact via hydrogen bonds, van der Waals forces, electrostatic interactions and so on. While in mechanical mixtures,
particles usually refers to atomic or molecular aggregates which are not bonded. Most of them are non-interactive and some may interact on
their surfaces electrostatically or via surface layers. In general, most colloidal dispersions can be considered as mechanical mixtures;
while real gases, liquid and solid solutions are chemical mixtures. However, to facilitate the analysis below, we define some mechanical
mixtures whose interactions on the surfaces can not be ignored or whose surface and volumetric heat content are strongly temperature-dependent
as quasi-chemical mixtures, i.e, mixtures in which irreversible thermodynamics can still be applied. In what presented below, analysis are
first performed on simple chemical mixtures, then followed by quasi-chemical mixtures, and finally on simple mechanical mixtures.

\subsection{Simple Chemical Mixtures}

For simplicity, it is assumed that the simple chemical mixture considered here is a binary solution and only thermal conduction and atomic or molecular
diffusion occur in it. Usually, the internal energy of a thermodynamic system is stored as the kinetic energy or
potential energy of atoms according to the their degrees of freedom. In ideal gases, the internal energy stored per degree of freedom per
atom is $\frac{1}{2} k_B T$ according to the equipartition theorem. Besides, it can also be stored as potential energy via van der Waals
interactions in real gases. In liquids, thermal energy is not only stored as the kinetic and potential energies of molecules but also stored as
potential energies via hydrogen bonds, electrostatic interactions and so on. In solids, the thermal energy is stored via the motion of phonons and free
electrons. At temperatures much lower than both the Debye and Fermi temperatures, the heat capacity of solid can be written as
$C_P = \gamma T + A' T^3$ where the contribution from electrons (phonons) is linear (cubic), $\gamma$ and $A'$ are material constants.$^{\cite{Kittel:1986}}$ Empirically, the molar heat capacity of substances in all three phases can be written as $C_P = a + b T + c T^{-2}$ where a, b, c are material
constants which varies for different phases and at different temperature ranges.$^{\cite{Brandes:1992}}$  Hence, the internal energy stored
in a substance per mole and the corresponding contribution to entropy per mole can be written as $\int_0^T C_P(\theta) d \theta$
and $\int_0^T \frac{C_P(\theta)}{\theta} d \theta$, respectively.

Now consider the mixing process of a ideal binary solution of components A and B at constant temperature T and pressure P. Let $u_i$, $s_i$, $v_i$, ($i=A, \, B,$) be the internal energy, entropy and volume per atom of components A and B before mixing; $n_i$ ($i=A, \, B,$) be the atom number
of A and B; $G'$, $U'$, $S'$, $V'$ be the total Gibbs free energy, internal energy, entropy and volume of the solution after the mixing. Then in this binary solution,
\begin{eqnarray}
U' &=& n_A u_A + n_B u_B \\
V' &=& n_A v_A + n_B v_B \\
S' &=& n_A s_A + n_B s_B + (n_A + n_B) \Delta S_{mix}
\end{eqnarray}
where there are no changes in the internal energy and volume since (1) the total energy is conserved and (2) the molar volume difference between species is neglected; the change in entropy arises from the contribution of the mixing entropy per atom $\Delta S_{mix}$. Usually, $\Delta S_{mix}= - k_B [c_A \, \ln(c_A) + c_B \, \ln(c_B)]+\Delta S^{ex}_{mix} $. In ideal solutions, the excessive entropy of mixing $\Delta S^{ex}_{mix}$ is zero. In nonregular solutions, $\Delta S^{ex}_{mix}=-(\frac{\partial \Delta G^{ex}_{mix}}{\partial T})|_{P,c_B}$, where $\Delta G^{ex}_{mix}$ is the excess free energy of mixing.$^{\cite{DeHoff:1993}}$ Hence, the total Gibbs free energy of the solution is
\begin{eqnarray}
G' &=& U' - TS' + PV' \nonumber \\
   &=& n_A (u_A-T s_A + P v_A) + n_B (u_B - T s_B + P v_B) - (n_A + n_B) T \Delta S_{mix}.
\end{eqnarray}
Convert it into the Gibbs free energy per unit volume by a multiplication of $\frac{\rho_0}{n_A+n_B}$, then
\begin{eqnarray}
G &=& c_A (U_A-T S_A + P V_A) + c_B (U_B - T S_B + p V_B) - \rho_0 T \Delta S_{mix} \nonumber \\
  &=& c_A G_A + c_B G_B - \rho_0 T \Delta S_{mix}                       \label{eq:11}
\end{eqnarray}
where $\rho_0$ is the number density of atoms per unit volume, $G_i$ $(i= A, \, B)$ are the Gibbs free energy of the pure A or B phase and $c=c_B$. This energy density function is usually preferred to study diffusion at constant T and P. However, in the binary solution considered here, there exist both
heat conduction and diffusion, then both T and c varies. Thus, in this function, the terms which is dependent on T must be shown explicitly.
In pure phases, $G_i = U_i-T S_i + P V_i$ $(i= A, \, B)$. Here, the first term ($U_i$) is the internal energy of certain reference
state and it is temperature dependent. Thus, $U_i$ can be written as $U^i_0 + \int_{T_0}^T C^i_P(\theta) d \theta$.
The third term ($P V_i$) arises from the mechanical work done on the system. It is temperature independent when
 thermal expansion is ignored. The second term ($-TS_i$) arises from the heat absorbed by the solution and it deserves a careful examination.
 Besides its explicit dependence on temperature, the entropy itself, written as $\int_0^T \frac{C_P(\theta)}{\theta} d \theta$ as shown above, is also temperature dependent and it increases as temperatures increases. Thus, this entropy can be written as
\begin{equation}
S_i = S^0_i + \int^T_{T_0} \frac{C^i_P}{\theta} d \theta \,\,\, (i= A, \, B), \label{eq:12}
\end{equation}
where $T_0$ and $S^0_i$ are the reference temperature and the corresponding entropy at the reference state $(T_0, \, P)$. Substitute
Eqn (\ref{eq:12}) into Eqn (\ref{eq:11}), then for a solution with both variant T and c but constant P, the free energy density function is
\begin{eqnarray}
G &=& c_A [U^A_0+\int_{T_0}^T C^A_P(\theta) d \theta-T S^0_A + P V_A] + c_B [U^B_0+\int_{T_0}^T C^B_P(\theta) d \theta - T S^0_B + P V_B] \nonumber \\
  & & \, - T (c_A \int^T_{T_0} \frac{C^A_P}{\theta} d \theta + c_B \int^T_{T_0} \frac{C^B_P}{\theta} d \theta )  - \rho_0 T  \Delta S_{mix} \\
  &=& c_A [U^A_0+\int_{T_0}^T C^A_P(\theta) d \theta] + c_B [U^B_0+\int_{T_0}^T C^B_P(\theta) d \theta] + P (c_A V_A + c_B V_B) \nonumber \\
  & & \, - T \{ c_A S^0_A + c_B S^0_B + \int^T_{T_0} \frac{c_A C^A_P + c_B C^B_P}{\theta} d \theta  + \rho_0 \Delta S_{mix} \}
\end{eqnarray}
As a result, the entropy density function for the ideal solution with both variant T and c is identified as
\begin{eqnarray}
S &=&  (1-c) S^0_A + c S^0_B + \int^T_{T_0} \frac{c_A C^A_P + c_B C^B_P}{\theta} d \theta + \rho_0 \Delta S_{mix} \nonumber  \\
  &=&  (1-c) S^0_A + c S^0_B + \int^T_{T_0} \frac{C_P(c,\theta)}{\theta} d \theta + \rho_0 \Delta S_{mix} \label{eq:13}
\end{eqnarray}
where the linear combination of the heat capacity is replaced by a more general expression of $C_P(c,T)$ in the second step.
From the above equation, it can be seen that the entropy density function for the ideal solution consists of a linear combination
of the entropies from the two pure phases, the thermal entropy owing to the heat transfer,
and the configurational entropy due to the mixing process. Note, the thermal and configurational entropies depends on
both temperature and composition. The entropy function itself is a state function and thus is independent of path. Then the
expression in Eqn (\ref{eq:13}) can be considered as being contributed from two processes. The first one is a mixing process
at temperature $T_0$ which leads to the linear combination of entropies of pure elements, i.e., the first two terms in Eqn
(\ref{eq:13}), and the mixing entropy $\rho_0 \Delta S_{mix}$. The second one is a heating process with the system being heated
from $T_0$ to $T$, which contributed to the third term $\int^T_{T_0} \frac{C_P(c,\theta)}{\theta} d \theta$. When the mixing entropy
is temperature-dependent, then the heating process also leads to an variation in the mixing entropy. Note that, as in the analyses
for the TE effects above, contribution from the gradient terms to the entropy function is also ignored here.

Next, consider a system described above with both non-equilibrium composition and temperature fields. Assume both
gradients of composition and temperature are moderate so that local thermodynamic equilibrium holds everywhere. Then according to
Eqn (\ref{eq:13}), the rate of change of the total entropy $\textbf{S}$ in the system is
\begin{eqnarray}
\frac{d\textbf{S}}{dt} &=& \int_V \frac{\partial S}{\partial t} dv \nonumber \\
 &=&  \int_V \left\{ [S^0_B - S^0_A + \int_0^T \frac{\frac{\partial C_P(c, \theta)}{\partial c}}{\theta} d \theta + \rho_0 \frac{\partial \Delta S_{mix}}{\partial c}] c_t + [ \frac{C_P(c, T)}{T} + \rho_0 \frac{\partial \Delta S_{mix}}{\partial T}] T_t \right\} dv \nonumber \\
 &=& \int_V \left\{ \Lambda' c_t + \Omega' C_p T_t \right\} dv, \label{eq:14}
\end{eqnarray}
where $\Lambda'$ and $\Omega'$ are introduced for brevity. The equation of conservation of mass are
\begin{eqnarray}
\frac{\partial c}{ \partial t} &=&  - \nabla \cdot \vec{J_c}. \label{eq:15}
\end{eqnarray}
Substitute Eqns (\ref{eq:3}) and (\ref{eq:15}) into Eqn(\ref{eq:14}), then
\begin{eqnarray}
\frac{d\textbf{S}}{dt} &=& \int_V \left\{ - \Lambda' \nabla \cdot \vec{J_c} - \Omega' \nabla \cdot \vec{J_q} \right\} dv \\
&=& \int_V \left\{ \vec{J_c} \cdot \nabla \Lambda' + \vec{J_q} \cdot \nabla \Omega' \right\} dv - \oint_A \left\{ \vec{J_c} \Lambda' + \vec{J_q} \Omega' \right\} \cdot \vec{n} da
\end{eqnarray}
where integration by parts is used in the second step. In the above formula, the first integrand in the volume integral is the local rate of
the entropy product and the second integrand in the surface integral is the local rate of change of the entropy exchanged with the surroundings.
Thus, to guarantee the local rate of entropy production to be positive, the fluxes need be defined as
\begin{eqnarray}
 \vec{J_c} &=& M_c(c,T) \nabla \Lambda' \\
 \vec{J_q} &=& M'_q(c,T) \nabla \Omega' \label{eq:16}
\end{eqnarray}
where $M_c(c,T)$ and $M'_q(c,T)$ are coefficients related to diffusivity and thermal conductivity. Here, the
mass flux equation is considered first. Expansion and simplification of this equation give
\begin{eqnarray}
 \vec{J_c} &=& M_c(c,T) \{ [ \int_0^T \frac{\frac{\partial^2 C_P(c, \theta)}{\partial c^2}}{\theta} d \theta + \rho_0 \frac{\partial^2 \Delta S_{mix}}{\partial c^2} ] \nabla c + [ \frac{\frac{\partial C_P(c, T)}{\partial c}}{T}  + \rho_0 \frac{\partial^2 \Delta S_{mix}}{\partial c \partial T } ] \nabla T \} \label{eq:21}\\
           &=& \frac{M_c(c,T)}{T} \left \{ - [ \rho_0 k_B T \frac{1}{c(1-c)} - \rho_0 T \frac{\partial^2 \Delta S^{ex}_{mix}}{\partial c^2} ] \nabla c  + [ \frac{\partial C_P(c, T)}{\partial c} + \rho_0 T \frac{\partial^2 \Delta S^{ex}_{mix}}{\partial c \partial T } ] \nabla T \right \} \\
           &=& -\rho_0 D [1- \frac{c(1-c)}{k_B } \frac{\partial^2 \Delta S^{ex}_{mix}}{\partial c^2} ] \nabla c + \frac{D}{k_B T} c(1-c)[ \frac{\partial C_P(c, T)}{\partial c}  + \rho_0 T \frac{\partial^2 \Delta S^{ex}_{mix}}{\partial c \partial T } ] \nabla T     \\
           &=& -\rho_0 D \nabla c - \rho_0 c(1-c) \left[ -\frac{D}{\rho_0 k_B T} \frac{\partial C_P(c, T)}{\partial c} - \frac{D}{k_B} \frac{\partial^2 \Delta S^{ex}_{mix}}{\partial c \partial T } \right] \nabla T  \;\;\; (in\; fluid\; phases)    \\
           &=& - \rho_0 D \nabla c - \frac{D}{k_B T} c(1-c) \left[ - T \frac{\partial C_P(c, T)}{\partial c} - \rho_0 T^2 \frac{\partial^2 \Delta S^{ex}_{mix}}{\partial c \partial T }  \right] \frac{\nabla T}{T}   \;\;\; (in\; solid\; phases)
\end{eqnarray}
At the second step, $\frac{\partial^2 C_P(c, \theta)}{\partial c^2}$ is assumed negligible and the expression of $\Delta S_{mix}$ is substituted; at the third step, $\frac{M_c(c,T)}{T}$ is taken as $\frac{D}{k_B T} c(1-c)$; at the last two steps, $\frac{\partial^2 \Delta S^{ex}_{mix}}{\partial c^2}$ associated with $\nabla c$ is assumed negligible. Then, the Soret coefficient in liquid phases and the heat of transport in solid phases are identified as
\begin{eqnarray}
S_T=-\frac{1}{\rho_0 k_B T} \frac{\partial C_P}{\partial c} - \frac{1}{k_B} \frac{\partial^2 \Delta S^{ex}_{mix}}{\partial c \partial T } \label{eq:22} \\
Q^*=- T \frac{\partial C_P}{\partial c} - \rho_0 T^2 \frac{\partial^2 \Delta S^{ex}_{mix}}{\partial c \partial T }, \label{eq:23}
\end{eqnarray}
respectively. Thus, the governing equation for the composition field is
\begin{eqnarray}
 \frac{\partial c}{\partial t} &=& -\nabla \cdot \vec{J_c} \\
           &=& \nabla \cdot \left \{ \rho_0 D(c,T)  [ \nabla c + c(1-c) S_T(c,T) \nabla T] \right \} \;\;\; (in\; fluid\; phases); \label{eq:19}\\
           &=& \nabla \cdot \left \{ D_0 e^{E_a/k_B T}  [\rho_0 \nabla c + \frac{c(1-c)}{k_B T} Q^* \frac{\nabla T}{T}] \right \} \;\;\; (in\; solid\; phases); \label{eq:20}
\end{eqnarray}
where the diffusivity $D(c,T)$ in liquid phase can be approximately assumed to be a constant if its field-dependence is not strong; while
in the solid phase the diffusivity must be explicitly expressed as $D_0 e^{E_a/k_B T}$ since a non-homogeneous distribution of temperature
is likely to significantly influence the jumping rate of atoms in the crystal lattice.

For the heat flux Eqn (\ref{eq:16}), it can be written as
\begin{eqnarray}
 \vec{J_q} &=&  M'_q(c,T) \nabla( \frac{1}{T} + \rho_0 \frac{1}{C_p} \frac{\partial \Delta S_{mix}}{\partial T} ) \nonumber \\
         &=& - \frac{M'_q(c,T)}{T^2} ( 1 + \rho_0 T^2 \frac{1}{C^2_p} \frac{\partial C_p}{\partial T} \frac{\partial \Delta S^{ex}_{mix}}{\partial T} - \rho_0 T^2 \frac{1}{C_p} \frac{\partial^2 \Delta S^{ex}_{mix}}{\partial T^2} )\nabla T  \nonumber \\
         & & + M'_q(c,T) \rho_0 (-  \frac{1}{C_p^2} \frac{\partial C_p}{\partial c} \frac{\partial \Delta S^{ex}_{mix}}{\partial T} +  \frac{1}{C_p} \frac{\partial^2 \Delta S^{ex}_{mix}}{\partial T \partial c}  )\nabla c \nonumber \\
         &=& - M_q(c,T)  \nabla T, \label{eq:17}
\end{eqnarray}
where at the second step, the expression of $\Delta S_{mix}$ is substituted and the Dufour effect is explicitly shown; at the third step, the heat conductivity $M_q(c,T)= \frac{M'_q(c,T)}{T^2}$ and assuming $\Delta S^{ex}_{mix}=0$ so that the Dufour effect can be ignored. Then using Eqn (\ref{eq:17}), the governing equation for the temperature field is
\begin{equation}
\frac{\partial T}{ \partial t} =   \frac{1}{C_P} \nabla \cdot [M_q(c,T)  \nabla T]. \label{eq:18}
\end{equation}
Thus, in fluid phases, the governing equations are Eqns (\ref{eq:19}) and (\ref{eq:18}); in solid phases, they are Eqns (\ref{eq:20}) and (\ref{eq:18}).

In the following discussions, the physical contents of the Soret coefficient and the heat of transport are further analysed. As shown above,
both of them have two contributions: the first one is a universal term and it is proportional to $-\frac{\partial C_P}{\partial c}$; the second one
 is related to the excessive entropy of mixing and it only exists in nonregular solutions. The mixture considered here is
binary one with components being atoms or molecules A and B, and c is the composition or concentration of species B. For simplicity, assume the mixture is a regular solution. Furthermore, assume interactions among
atoms or molecules can be ignored in these mixtures (e.g, ideal gases), then heat can not be stored as the potential energies via interactions and
the contribution to heat capacity arises from the heat stored in each atom or molecule. As a result, the molar heat capacity is a
linear combination of heat capacity of single A and B atom or molecule ($C^A_P$ and $C^B_P$), i.e, $C_P = (1-c) C^A_P + c C^B_P$.
Then, for these mixtures, both $S_T$ and $Q^* \propto (C^A_P-C^B_P)$. If $C^A_P > C^B_P$, then B atoms or molecules move against the temperature
gradient to the cooler area and vice versa. In both cases, the atoms or molecules with large molar heat capacities move to the
hotter area.

Consider binary ideal gases with both molecules being monoatomic or diatomic, since
$C^A_P=C^B_P=\frac{3k_B}{2}$ or $\frac{5k_B}{2}$, then $S_T=0$. In these gases, the molecules
are thermally identical particles. The separation of them will not cause a flow of thermal energy, i.e, the Dufour effect. Then, it is
reasonable that a flow of thermal energy will not cause a mass flux for separation according to Onsager's relationship. However, for binary
ideal gases mixtures with one molecule being monoatomic and the other being diatomic, let $C^A_P=\frac{5k_B}{2}$ and $C^B_P=\frac{3k_B}{2}$,
then $S_T=\frac{1}{T}$. Thus, the monoatomic B molecules move to the area with
lower temperatures and the diatomic A molecules move to the area with higher temperatures. Also in solder joints, it was found that Sn atoms
move to the hot side under the influence of thermomigration in both eutectic SnPb alloy ($80.9-111.2^oC$ and $1000^oC/cm$)$^{\cite{Chuang:2006}}$ and
SnAg3.5 alloy ($134.3-154.3^oC$ and $2829^oC/cm$).$^{\cite{Hsiao:2009}}$
Since no sign change of $Q*$ of Sn was observed, then the nonlinear contribution to the molar heat capacity can be assume small (to be explained in the
next paragraph) and $C_P$ is approximately a linear combination of the heat capacities per atom basis of the two species. The calculated molar heat capacities
$(\frac{J}{K \, mol})$ of the two species at the average temperatures are 27.21 for Pb and 28.38 for Sn ($100^oC$); and 25.74 for Ag and 29.18
for Sn ($144.3^oC$).$^{\cite{Brandes:1992}}$ In both alloys, Sn atoms have a larger molar heat capacity. Thus, according to our analysis they will move to the area
with higher temperatures, which is consistent with the above experimental observations.

Next, assume interactions among atoms or molecules can not be ignored (e.g., real gases, most liquids and solids), then the
heat can be stored as the potential energies due to these interactions (e.g., van der Waals interactions among gas molecules,
hydrogen bonds in liquids). Then, the molar heat capacity is no longer a linear combination of the molar heat capacities of single A and B
atom or molecule. When the nonlinear contribution, e.g., a quadratic contribution $\rho_0 c (1-c) C^{AB}_P$, results in a maximum or minimum in
the middle for some mixtures, then $\frac{\partial C_P}{\partial c}$ has a change of sign as c varies from 0 to 1. As a result, $S_T$ or $Q^*$ will
change sign as the concentration or composition varies. In simple liquid mixtures, it was observed that the Soret coefficient will change sign as the concentration of components varies.$^{\cite{Carlos:2005}}$ If $\frac{\partial C_P}{\partial c}>0$ (i.e, the segregation of B atoms resulting in an increase in
the molar heat capacity of the local phase), then $S_T<0$ and B atoms move to the area with higher temperatures which thus leads to a rich B
phase there; and vice versa. Moreover, since the heat capacity itself strongly depends on temperature T, then both $S_T$ and $Q^*$ have strong
dependencies on temperature. In solids, the heat of transport is usually shown by experiments to have a strong dependence on temperature.

In metallic solid phases, diffusion of vacancies is usually found to be subject to the influence of a temperature gradient. Since the concentration
of vacancies is usually around $10^{-4}$ at the room temperature, then the system can be considered as an ideal or regular solution. Hence,
the heat of transport of vacancies $Q^*_v$ is $- T \frac{\partial C_P}{\partial c}$, which can be rewritten as
\begin{eqnarray}
 Q^*_v &=& - T \frac{\partial}{\partial c}[\frac{\partial }{\partial T}\int_0^T C_P d\theta] \nonumber \\
       &=& - T \frac{\partial}{\partial T}[N_0 \frac{\partial }{\partial N_v}\int_0^T C_P d\theta] = - T \frac{\partial}{\partial T}[N_0 \frac{\partial }{\partial N_v} H(N_v, T) ] \nonumber \\
       &=& - N_0 T \frac{\partial h_v(T) }{\partial T}
\end{eqnarray}
where $N_0$ is the number density per unit volume and $N_v=N_0c$ is the number of vacancies per unit volume, $h_v= \frac{\partial }{\partial N_v} H(N_v, T)$ is the enthalpy of the formation of an vacancy. At the steady state, then $Q^*_v$ can be approximated as
\begin{eqnarray}
 Q^*_v = - N_0 T \frac{ h_v(T_h) - h_v(T_l) }{T_h-T_l},
\end{eqnarray}
which is proportional to the difference in the enthalpy of the formation of an vacancy at the hot and cold sides. If $h_v(T_h)>h_v(T_l)$, then $Q^*_v$
is negative and vacancies diffuse from the hold side to the cold side and vice versa. As argued previously, thermomigration arises from the conversion
 of heat into the chemical energy of the system. When an vacancy diffuse from the hot side to the cold side, according to the law of conservation of energy,
 the amount of heat converted into the chemical energy is $h_v(T_h)-h_v(T_l)$. Usually, vacancies which are produced at the hot side migrate to the cold
 side and annihilate there. When each one of them is annihilated, the heat released in the cold side is $h_v(T_l)$. At the same time, annihilation of an vacancy leads to the dissipation of the chemical free energy and it is $h_v(T_h)-h_v(T_l)$. As a result, the total amount of heat conducted by the final
 annihilation of an vacancy at the cold side is $h_v(T_h)$.

However, in nonregular solutions the second contribution due to $\frac{\partial^2 \Delta S^{ex}_{mix}}{\partial c \partial T }$ needs be taken
into account since this term is no longer zero. In regular solution models, the excessive entropy of mixing is usually assumed zero and the
excessive heat of mixing $\Delta H^{ex}_{mix}$ assumed temperature-independent. For examples, the simplest form of $\Delta H^{ex}_{mix}$
can be written as $a c(1-c)$ where a is a material constant. Thus, the excessive free energy of mixing
$\Delta G^{ex}_{mix}=\Delta H^{ex}_{mix}-T\Delta S^{ex}_{mix}=ac(1-c)$. But in nonregular solutions, the excessive free energy of mixing
is usually temperature dependent and its simplest form can be $ac(1-c)(1+\frac{b}{T})$ where b is also a material constant. As a result,
the excessive entropy of mixing is $\Delta S^{ex}_{mix}= -(\frac{\partial \Delta G^{ex}_{mix}}{\partial T})|_{P,c}=\frac{ab}{T^2}c(1-c)$,
which depends on both temperature and composition. Hence, the second contribution due to $\Delta S^{ex}_{mix}$ is non-zero. In this case,
determination of $S_T$ and $Q^*$ is more complex due to the temperature dependence of the heat of mixing.
That is, the mixing process itself produces heat and the amounts of heat produced are different at the areas with lower or higher temperatures. Thus,
this contribution must be taken into account.

Finally, to conclude this subsection, the physical content of the driving force due to a thermal gradient is discussed, which is
the key to understand the mechanism of TDMT. According to Eqn (\ref{eq:21}), the driving force $F_T$ due to the temperature gradient can be found as
\begin{equation}
F_T = [\rho_0 \frac{\frac{\partial C_P(c, T)}{\partial c}}{T}  + \rho_0 \frac{\partial^2 \Delta S^{ex}_{mix}}{\partial c \partial T }] \nabla T,
\end{equation}
which is the dominant term of $\nabla \frac{\partial Q(c,T)}{\partial c}$ assuming $\frac{\partial^2 C_P}{\partial c^2}$ negligible and using $\Delta S^{ex}_{mix} =0 $ in regular solutions. The chemical driving force is $F_{ch}=-\nabla \frac{\partial f_{ch}}{\partial c}$. Thus, the total driving
force F for atomic diffusion can be approximated by
\begin{equation}
F_{ch} + F_T = -\nabla \frac{\partial f_{ch}}{\partial c} + \nabla \frac{\partial Q(c,T)}{\partial c}.
\end{equation}
Consider the steady state of TDMT where the mass flux is zero. That is, the mass flux due to TDMT and chemical driving force lies in opposite directions and
cancel each other, which is similar to the limiting case during electromigration when the growth of intermediate phase is retarded. Then, there is a fixed compositional profile and a constant temperature gradient. Since the total mass flux is zero, then the total driving force for diffusion is zero. Thus, the chemical free energy is no longer consumed. However, the chemical driving force arising from the compositional gradient will continuously leads to the consumption of
chemical free energy. Then, the other driving force must convert some energy in another form at the same rate into the chemical free energy in order to
maintain a steady state. As shown by the formula of $F_T$, this energy has to be heat. Thus, it is reasonable to argue that TDMT arises from the conversion
of heat into the chemical energy of the system. This is similar to the mechanism of electromigration which arises from the conversion of the electrical work
into the chemical energy of the system.

In solids, heat is conducted by the flow of electrons and phonons. At the presence of a temperature gradient,
both electrons and phonons flow from the hot side to the cool side. Usually, the momentum of electrons consists of the Fermi momentum and the drift momentum. The former determines the translational energy of electrons, i.e, the heat stored by electrons; while the latter arises from the externally applied electric
field and thus is related to the electrical work. Under the influence of a temperature gradient, the
distribution of Fermi momentum of electrons is not at equilibrium. The average Fermi momentum is larger (smaller) at areas with higher (lower)
temperatures, thus there is a flow of electrons from the hot area to the cold area. As a result, the Fermi momentum is transferred to diffusing
atoms during the scattering effects of these two. This aids the jumping of atoms on lattice sites and then help change the local configuration of atoms in
the local volume element, which is related to its local chemical energy density. Since Fermi momentum is related to the heat
stored by electrons, then the heat of electrons is converted into the chemical energy of atoms during the process of thermomigration. Consequently,
in solid phase it is reasonable to argue that the conversion of heat into the chemical energy of the system is the cause of thermomigration. And in general,
the cause of the TDMT in all three phases of matter since they share the same physical ground.

In electromigration, the momentum transfer between electrons and atoms was once used
to argue against the linear dependence between the effective charge number and the current density. Because the Fermi momentum is orders of
magnitudes larger than the drift momentum, thus the transfer of Fermi momentum, which is independent on current density, to electrons is dominant.
Consequently, the effective charge number is independent on current density. However, our views are different on this matter.
To separate the effects of electromigration and thermomigration, it is necessary to assume that during the process of
electromigration, heat is quickly conducted away and thus the system is maintained thermally uniform. As a result, the distribution of
Fermi momentum is at equilibrium and the average Fermi momentum of electrons is everywhere the same. However, due to the externally applied
electric field, electrons obtain drift momentum. They lose or partially lose this momentum during the collision between them and atoms, and
then regain momentum via re-accelerations in the electric field. Thus, during the process of electromigration, the drift momentum is
transferred to the diffusing atoms and as a result the electrical work is converted into the chemical energy of the system. In brief, we
argue thermomigration arises or partially arises from the transfer of the Fermi momentum of electrons to diffusing atoms, the other contributions are
from phonons; while electromigration arises from the the transfer of the drift momentum of electron to diffusing atoms. Thus the former is related
to the conversion of heat into the chemical energy of the system while the latter the conversion of the electrical work. This is believed to be the substantial difference between thermomigration and electromigration.

\subsection{Quasi-chemical Mixtures}

In the former subsection, TDMT is discussed in simple chemical mixtures which only involves atoms and molecules. In this
subsection, particles are not only atoms and molecules but also colloidal particles, i.e, aggregates of atoms or molecules. Usually,
the diameters of colloidal particles are around 1-1000 nanometers. Here, we define mixtures with colloidal particles whose diameters are round 10
nanometers (1 micrometers and above) as quasi-chemical (simple mechanical) mixtures. Thus, an average particle in quasi-chemical mixtures includes around
100 atoms while that in simple mechanical mixtures includes 10 million atoms. Then, in quasi-chemical mixtures, colloidal particles can be
treated as being at thermal equilibrium locally with surrounding solvent particles. However, in simple mechanical mixtures, colloidal particles,
owing to the large number of atoms they contained, are usually not at thermal equilibrium locally and heat is usually conducted inward from the surface
into the center of the colloidal particles. In both analyses by Brock and Yalamov et. al, heat conduction within the colloidal particles are
considered.$^{\cite{Derjaguin:1965,Brock:1962}}$ As a result, we argue that the principles of irreversible thermodynamics, which is based on local
thermodynamic equilibrium, can still be applied to quasi-chemical mixtures but not to simple mechanical mixtures.

In quasi-chemical mixtures, heat can be stored via the interactions between the atoms or molecules on the surface of the colloidal particles with those in the solvent surrounding them. Moreover, it can also be stored as the kinetic or potential energies of the atoms or molecules within the colloidal particles. The former
is a surface contribution and the latter a volumetric contribution. Thus, for an average colloidal particle with an average diameter r, its total
heat capacity $C^{*}_p$ is written as
\begin{equation}
C^{*}_p = \rho^s_0 c^{s}_p 4 \pi r^2 + \rho^{v}_0 c^{v}_p \frac{3}{4} \pi r^3,
\end{equation}
where $\rho^s_0$ ($\rho^v_0$) is the number density of atoms per unit surface (volume) of the particle, $c^{s}_p$ ($c^{v}_p$) is the heat capacity per
 atom on the surface (whthin the volume) of the particle. As a result, the heat capacity per unit volume of the quasi-chemical mixture is
\begin{eqnarray}
C_p &=& \frac{N C^{*}_p + (V-N 3 \pi r^3/4)\rho'_0 c'_p}{V} \nonumber \\
    &=& \rho'_0 c'_p + c \rho_0^s c_p^{s} 4 \pi r^2 + c (\rho_0^v c_p^{v}-\rho'_0 c'_p) \frac{3}{4} \pi r^3
\end{eqnarray}
where $c=N/V$ is the concentration of particles, $\rho'_0$ is the number density of atoms per volume of the solvent and $c'_p$ is the heat capacity per atom of the solvent. As a result, using Eqns (\ref{eq:22}) and (\ref{eq:23}), in quasi-chemical mixtures we have
\begin{eqnarray}
S_T= - \frac{1}{\rho_0 k_B T} [\rho_0^s c_p^{s} 4 \pi r^2 + (\rho_0^v c_p^{v}-\rho'_0 c'_p) \frac{3}{4} \pi r^3] - \frac{1}{k_B} \frac{\partial^2 \Delta S^{ex}_{mix}}{\partial c \partial T } \\
Q^*= - T [\rho_0^s c_p^{s} 4 \pi r^2 + (\rho_0^v c_p^{v}-\rho'_0 c'_p) \frac{3}{4} \pi r^3] - \rho_0 T^2 \frac{\partial^2 \Delta S^{ex}_{mix}}{\partial c \partial T },
\end{eqnarray}
where $\rho_0$ is the average number density of solvent atoms and solute particles per unit volume. As can be seen, in quasi-chemical mixtures $S_T$ and $Q^*$
consists of three contributions. The first term arises from the heat capacity contributed from the atoms at the surface of the solute particles and it
 is proportional to the total area of the surface of the particle. The second term arises from the difference between the heat capacity contributed from
 the solute particle and the capacity of solvent atoms which occupy the same space of the solute particle. Hence, the second term is proportional to the volume of the particle. The third term arises from the excessive mixing entropy as discussed previously. Generally speaking, when it is difficult to achieve thermal equilibrium between the bulk of the particle and the solvent, and furthermore when the excessive mixing entropy is negligible, then the first term is the dominant term. Thus, the Seebeck coefficient is usually found to be proportional to the total area of the particle surface.$^{\cite{Braun:2006}}$

\subsection{Mechanical Mixtures}

Mixtures with colloidal particles whose diameters are around or larger than 1 micrometers are defined here as simple mechanical mixtures. In this
case, due to the large size and the large number (10 million) of atoms within the bulk of the suspended particles, thermal and chemical equilibriums can only
be achieved at the surface of the particles. As a whole, the particles and the surrounding media are usually not at local thermodynamic equilibrium and there exists a radial temperature gradient within the particles. More importantly, as compared to the factor affecting TDMT to be discussed below, the surface
contribution to entropy and the excessive mixing entropy can be considered negligible. In subsection 3.1, it is argued that TDMT in simple chemical mixtures
arises from the conversion of heat into the chemical energy of the system. However, in simple mechanical mixtures, atoms or molecules from the hot side
carrier more kinetic energy than those from the cold side. Thus, collisions of these atoms or molecules at the surface of the suspended
particles results in a net driving force from the hot side to the cold side. In this case, the suspended particles gained kinetic energies from the kinetic
 energies of the surrounding atoms or molecules. The former is related to the mechanical energy of the suspended particles while the latter
 the thermal energy of the solvent atoms or molecules. Thus, in this case, heat of the surrounding solvents is converted into the mechanical energy of the suspended particles. Though the chemical energy of the atoms at the surface of the suspended particles will vary when the particles are driven from the hot
 side to the cold side. However, compared to the variation in the mechanical energy of the particles, it can be considered negligible. In this case, PFVA is
 no longer applicable to TDMT in mechanical mixtures and the traditional hydrodynamic method need be used to determine the thermophoretic force.$^{\cite{Piazza:2008}}$

In this section, TDMT in simple chemical, quasi-chemical and mechanical mixtures are discussed. In simple and quasi-chemical mixtures, a temperature gradient
in an initially chemically uniform media results in a compositional gradient. Thus, heat is converted into the chemical free energy of the system. When considered from
the perspective of the variation of the total entropy, as shown in Eqn (\ref{eq:13}), the total entropy consists of the thermal entropy and the mixing entropy (configurational entropy). Since the chemical free energy is increased, then the configurational entropy itself is lowered. Moreover, the rate
of production of entropy is everywhere non-negative. Then, the thermal entropy of the system and its surroundings must increase and its increment must be
larger than the decrement in the configurational entropy to guarantee that the total entropy increases. In brief, the decrement of the configurational entropy
of the system is at the cost of a larger increment of the thermal entropy of the universe. The configurational entropy itself is a function of composition, and
as suggested by the name is related to the configuration of atoms or molecules within the volume element. The configurational entropy is a maximum when the
system is chemically uniform, i.e,, when the system is disordered to the largest extent. As TDMT leads to a compositional gradient within the system, the system becomes more ordered and the configurational entropy lowers. Hence, TDMT is actually a process which directly causes order to arise from disorder.
(Electromigration can achieve this effect as well.) Thus, it is believed TDMT could be related to the origin of life.$^{\cite{Wurger:2006}}$ According to the ``Soup" theory, order-from-disorder in ``primordial soup" is believed to be an important step towards abiogenesis. Compared with an electrical field, temperature gradients are very common near natural environments such as
hot springs, oceanic vents and so on. Furthermore, chemical reactions itself release heat into surroundings and produce temperature gradients around. It is
thus believed owing to these temperature gradients, TDMT leads to order-from-disorder and probably played an important role in abiogenesis. In the ``soup", TDMT leads to the separation of simple organic compounds near the hot and cold areas. Chemical reactions of these simple compounds results in complex organic
polymers and the heat released provides further driving forces for TDMT. Iteration of these processes leads to more orderliness in the soup and laid down
the foundation of abiogenesis. In a sense, the origin life is related to the lower of configurational entropy and the conversion of heat into the chemical
free energy by TDMT in a micro-organic system. When complex organic compounds become more ordered and group into certain units, their sizes continually grow
due to both chemical reactions and TDMT. Usually, large sizes results in instability. Moreover, chemical reactions occur on the surface of these units
results in backward temperature gradients within these units. Their directions are opposite to those of the outer temperature gradients which originally helped
the formation of these units. This could cause their inner cores to decompose and help form shell-like structures. These unstable structures tend to
break into pieces and each one of these pieces serves as a new site for the new order-from-disorder process. That is, units at the early stage of abiogenesis
probably accomplished self-production in pure mechanical ways. In brief, because TDMT leads to order-from-disorder, it is reasonable to believe it is related
to abiogeneses.

In previous two sections, both TDMT and TE have been analyzed using PFVA with entropy functionals. Start from the section below, PFVA with energy
functionals will be used to analyze irreversible processes and the conversion and dissipation of energies will be discussed.

\section{Conversion and Dissipation of Free Energies}

As mentioned in the introduction, complex irreversible processes consists of coupling effects between two or more simple irreversible processes.
During these processes, energies are not only dissipated but also converted from on form into another. For examples, electromigration (thermomigration)
arises from the conversion of electrical work (heat) into the chemical energy of the system; TE effects are related to the conversion between electrical
 energy and heat. To study the conversion and dissipation of free energies, consider a thermodynamic system which involves two physical fields $\psi$ and $\eta$. First, linear irreversible processes are analyses using Onsager's formulation by assuming the system is close to equilibrium. Then, nonlinear irreversible processes are analysed with PFVA as a general description of a system near equilibrium.

\subsection{Linear Irreversible Processes using Onsager's approach}

For linear irreversible processes, it is assumed that a given flux $\vec{J}_i$, instead of only depending on its conjugate driving force $\vec{X}_i$, depends on all driving forces in the system, i.e., $\vec{J}_i = \vec{J}_i(\vec{X}_1, \vec{X}_2, ...)$. When the system is close to equilibrium, the given flux can be expanded as $\vec{J}_i=\displaystyle{\sum_j} L_{ij} \vec{X}_j$ with the second and higher order terms ignored. For the coupling coefficients $L_{ij}$ ($i\neq j$), Onsager proposed that $L_{ij}=L_{ij}$ and a proof based on microscopic reversibility was given. This is known as Onsager's relations, which contributed to the proof of Thomson's second relation for TE effects. Latter, Casimir corrected these relations by considering that fields such as velocities and the magnetic field are associate with sign changes under the performance of time reversal. Experimental data were found to be in favor of the validity of Onsager's relations.$^{\cite{Miller:1960}}$ However, owing to a lack of macroscopic proof, they were questioned$^{\cite{Truesdell:1984}}$ and thus are advised to be regarded as ``postulates at the macroscopic level".$^{\cite{Jou:2010}}$ In the following discussions, Onsager's relations are analyzed. In the system considered here, let $E_\psi$ and $E_\eta$ denote the total free energies associated with the two physical fields. The corresponding free energy densities are $\epsilon_\psi$ and $\epsilon_\eta$. Then, according to Onsager's approach, the fluxes associate with the two physical fields $\psi$ and $\eta$ can be written as
\begin{eqnarray}
\vec{J}_{\psi} &=& -L_{\psi\psi} \nabla \mu_{\psi} -L_{\psi\eta} \nabla \mu_{\eta}  \label{eq:31}\\
\vec{J}_{\eta} &=& -L_{\eta\eta} \nabla \mu_{\eta} -L_{\eta\psi} \nabla \mu_{\psi}, \label{eq:32}
\end{eqnarray}
where $\mu_{\psi}=\frac{\partial \epsilon_\psi}{\partial \psi}$ and $\mu_{\eta}=\frac{\partial \epsilon_\eta}{\partial \eta}$. Here, the driving forces are defined as $-\nabla \mu_{\psi}$ and $-\nabla \mu_{\eta}$. Note that in Onsager's approach, driving forces serve as independent variables and their forms remain the same as those in simple irreversible processes. Consequently, an important assumption implicit in Onsager's approach is that the energies either have no coupling dependence or their coupling dependence can be ignored. That is, $\epsilon_\psi = \epsilon_\psi(\psi)$ which is either independent of $\eta$ or its dependence on $\eta$ can be ignored; and likewise for $\epsilon_\eta$. Otherwise, this interdependence of the energies will lead to modifications of the driving forces and thus they are no longer the same as those in simple irreversible processes.

Now, assume initially in the system the distribution of the field $\eta$ is uniform and an externally applied field associated with $\psi$ results in a gradient in $\psi$ instantaneously. Due to the complex irreversible process between $\psi$ and $\eta$, this gradient eventually leads to a gradient in $\eta$. Consider the rate of change of the energy $E_{\psi}$, then we have
\begin{eqnarray}
\frac{dE_{\psi}}{dt} &=& \int_V \frac{\partial \epsilon_{\psi}}{\partial \psi} \frac{\partial \psi}{\partial t} dv  \nonumber \\
 &=& - \int_V (\frac{\partial \epsilon_{\psi}}{\partial \psi} \nabla \cdot \vec{J}_{\psi}) dv \nonumber \\
 &=& - \oint_A \frac{\partial \epsilon_{\psi}}{\partial \psi} \vec{J}_{\psi} \cdot \vec{n} da - \int_V L_{\psi \psi} \nabla \mu_{\psi} \cdot \nabla \mu_{\psi} dv -  \int_V L_{\psi \eta} \nabla \mu_{\psi} \cdot \nabla \mu_{\eta} dv, \label{eq:25}
\end{eqnarray}
where at the second step, the conservation law $\frac{\partial \psi}{\partial t }=-\nabla \cdot \vec{J}_{\psi}$ is used; at the last step, the first term is the energy flow across the boundary, the second term guarantees the energy decreases as time evolves in an isolated system without any coupling effects, the physical content of the third term needs careful examination. The driving forces are given by the negative gradients of $\mu_{\psi}$ and $\mu_{\eta}$. According to the third term, if the two driving forces lies in the same direction (opposite directions), then the rate of the consumption the total free energy $E_\psi$ is enhanced (retarded). Recall the polarity effect of electromigration during the intermediate phase growth . When the
chemical driving force and the effective driving force for electromigration lies in the same direction, the growth rate of the intermediate phase is enhanced and so is the rate of the consumption of the chemical free energy, and vice versa. It has been argued that electromigration reflects the conversion of the work by the electrical field into the chemical energy of the system. Then analogously, the third term here also reflects
the conversion between different energies, i.e., either the conversion of $E_{\psi}$ into $E_{\eta}$ or the conversion of $E_{\eta}$ into $E_{\psi}$.  When the driving forces lie in the same direction, then the third term is negative. In this case, the rate of consumption of $E_\psi$ is enhanced and an extra amount of $E_{\psi}$ should be converted into $E_{\eta}$ due to the coupling effect. On contrast, in the other case when the driving forces lie in the opposite directions, then the third term is positive. As a result, the rate of consumption of $E_\psi$ is retarded and an extra amount of $E_{\eta}$ should be converted into $E_{\psi}$ due to the coupling effect. That is, the third term on the r.h.s of Eqn (\ref{eq:25}) represents the conversion of energies. Note that, in the above discussions, $-\nabla \mu_{\psi}$ in the third term is the driving force due to the externally applied field $\psi$; $-\nabla \mu_{\eta}$ is the resulting driving force due to the coupling effect. That is, the former is the cause while the latter is the response. For example, in the Thomson's effect, the electrical field due to the externally applied current is the cause while the temperature gradient is the responding driving force.

Next, consider the rate of change of energy $E_{\eta}$ analogously and it is given by
\begin{eqnarray}
\frac{dE_{\eta}}{dt'} = - \oint_A \frac{\partial \epsilon_{\eta}}{\partial \eta} \vec{J}_{\eta} \cdot \vec{n} da - \int_V L_{\eta \eta} \nabla \mu_{\eta} \cdot \nabla \mu_{\eta} dv -  \int_V L_{\eta \psi} \nabla \mu_{\eta} \cdot \nabla \mu_{\psi} dv. \label{eq:26}
\end{eqnarray}
Its expression is similar to Eqn (\ref{eq:25}) but with the subscripts $\psi$ and $\eta$ switched. Note that, in this equation, as $t'$ evolves the energy $E_{\eta}$ also decreases as indicated by the second term, because its contribution to the rate of change of energy $E_{\eta}$ is usually major and that of the third term is minor. However, this is inconsistent with the situation in the system discussed above. When the gradient in $\psi$ results in a gradient in $\eta$ and eventually the system reaches a steady state, the energy $E_{\eta}$ in fact increases as time evolves. Thus, to enable the discussion of the variation of energies in the same system, a time reversal $t=-t'$ needs be performed for the above Eqn (\ref{eq:26}) so that when $E_\psi$ decreases with time t, $E_\eta$ increases. After the time reversal, the portion related to the conversion of energies is
\begin{equation}
\frac{dE^{cv}_{\eta}}{dt} =  \int_V L_{\eta \psi} \nabla \mu_{\eta} \cdot \nabla \mu_{\psi} dv. \label{eq:27}
\end{equation}
Owing to the law of conservation of energy, the rate of change of the two energies due to the conversion of energies must sum up to zero. That is, at each instant during the conversion of energies, the gain or loss in $E_\psi$ must equal the loss or gain in $E_\eta$ in magnitude. Consequently,
\begin{equation}
\frac{dE^{cv}_{\psi}}{dt} + \frac{dE^{cv}_{\eta}}{dt} = - \int_V (L_{\psi \eta}-L_{\eta \psi}) \nabla \mu_{\eta} \cdot \nabla \mu_{\psi} dv \equiv 0.
\end{equation}
Here, the volume can be chosen arbitrarily so that $L_{\psi \eta}=L_{\eta \psi}$ directly follows. Note, this is the Onsager's relation when the complex irreversible process involves only two physical fields. Arguments can be presented analogously for complex irreversible processes involving multiple physical fields. When velocities, magnetic fields and so on are present in the driving forces, there is a sign change in the above time reversal and thus
will directly lead to the Casimir correction.

In Eqn (\ref{eq:25}), it is mentioned that the second term guarantees the free energy for a simple irreversible process decreases as time evolves in an isolated system. Simple irreversible processes usually directly lead to the dissipation of free energy. Thus the second term in fact reflects the dissipation of energy $E_{\psi}$. Usually, most simple irreversible processes dissipate energy as heat. For examples, electrical forces dissipate electrical energies as heat during the Joule heating and chemical driving forces also dissipate chemical free energies as heat during the diffusion process.

Then, as a result, Eqn (\ref{eq:25}) can be rewritten as
\begin{eqnarray}
\frac{dE_{\psi}}{dt} &=& \frac{dE_{\psi}^{cd}}{dt} + \frac{dE_{\psi}^{ds}}{dt} + \frac{dE_{\psi}^{cv}}{dt}, \label{eq:28} \\
    && \frac{dE_{\psi}^{cd}}{dt} = - \oint_A \frac{\partial E_{\psi}}{\partial \psi} \vec{J}_{\psi} \cdot \vec{n} da \\
    && \frac{dE_{\psi}^{ds}}{dt} = - \frac{dQ}{dt} = - \int_V L_{\psi \psi} \nabla \mu_{\psi} \cdot \nabla \mu_{\psi} dv \label{eq:29} \\ 
  && \frac{dE_{\psi}^{cv}}{dt}= - \frac{dE_{\eta}^{cv}}{dt} = - \int_V L_{\psi \eta} \nabla \mu_{\psi} \cdot \nabla \mu_{\eta} dv; \label{eq:30}  
\end{eqnarray}
where $\frac{dE_{\psi}^{cd}}{dt}$ represents the energy conducted by the flux $\vec{J}_{\psi}$, $\frac{dE_{\psi}^{ds}}{dt}$ represents the energy dissipated which is in the form of heat, and $\frac{dE_{\eta}^{cv}}{dt}$ represents the energy converted into $E_{\eta}$. That is, the energy $E_{\psi}$ can be
conducted, dissipated and converted. This is essentially the law of conservation of energy or the first law of thermodynamics. This law was discovered around one and half centuries ago when steam engines greatly propelled the industrial revolution. Thus, to
a large extent, the original statements of this law is closely related to the performance of steam engines. By steam engines, heat is converted into mechanical work for industrial use. Thus, as the first law states, ``the increase in internal energy of a closed system is equal to the heat supplied to the system minus work done by it". If $E_\psi$ is taken as the internal energy of the system, then $dE_\psi$ is the increase of internal energy; $dE^{cd}$ is the heat supplied and transferred via the boundaries into the system by the steam; $dE_{\eta}^{cv}$ is the work done by the system, i.e., the amount of internal energy converted into the mechanical work $E_{\eta}$. The dissipation of the internal energy is not explicitly shown in the original statements because at early times, the conduction of heat from within the engine into the environments, the internal friction of the steams and the energy dissipated during the phase transformation between steams and liquid water are not major concerns. Moreover, the heat supplied can also be considered as the net heat supplied, i.e., the heat flowed into the system minus the dissipation of the internal energy which is in the form of heat. Nowadays, the law of conservation of energy found a much broader application, which not only involves heat engines but also diffusion problems, phase transformations, chemical reactions, mechanical failures and so on. When complex irreversible processes are considered, energies are conducted, dissipated and converted. Thus, it is more helpful to show the dissipation and conversion of energies explicitly as shown in Eqn (\ref{eq:28}). More importantly, the dissipation and conversion of energies can also be numerically calculated using the forms shown in Eqns (\ref{eq:29}) and (\ref{eq:30}). Note that, in the discussions here, the dissipation (conversion) of energy is defined to be the transformation of one energy into heat (another energy) which is useless (useful).

Close examination of Eqns (\ref{eq:29}) and (\ref{eq:30})shows that the dissipation of energy $E_{\psi}$ is determined by the inner product of
the driving force $- \nabla E_\psi$ and its conjugate portion of the flux, $-L_{\psi \psi} \nabla \mu_{\psi}$; on contrast, the conversion of energy $E_{\psi}$ into $E_{\eta}$ is determined by the inner product of the driving force $- \nabla E_\psi$ and the coupling portion of the flux, $-L_{\psi \eta} \nabla \mu_{\eta}$. Using Onsager's relations, the latter can also be interpreted as the inner product between the other driving force $- \nabla E_\eta$ and the coupling portion of the other flux, $-L_{\eta \psi} \nabla \mu_{\psi}$. Thus, the inner product between the driving force and its conjugate portion of the corresponding flux represents the rate of dissipation of energy; while the inner product between the driving force and the coupling portion of the corresponding flux represents the rate of conversion of energy. This is actually nothing new. Fluxes measures the rate of change of energy, mass and so on, thus they can be considered as generalized velocities. In classic mechanics, the inner product between forces and velocities are usually defined as the rate of work, which is in fact either the dissipation or the conversion of the kinetic energy. For examples, consider a spring-mass system place on a smooth desktop. When the mass is displaced from the equilibrium position, the product of the restoring force and the velocity of the mass either represent the
conversion of the kinetic energy of the mass into the elastic energy of the spring or vice versa. In a simple diffusion (conduction) process, the inner product between the chemical (electrical) driving force and the mass (electron) flux represents the dissipation of the chemical free (electrical) energy
into heat. For a complex irreversible process such as electromigration, the mass flux is $\vec{J}_c = M_0 c(1-c) [ - \nabla \mu_c + e Z_e \vec{E} ]$ where $\mu_c=\frac{\partial f_{ch}}{\partial c}$ and $Z_e$ is the effective charge number(usually negative).$^{\cite{Zhou:2011}}$ Thus, the rate of change of the chemical free energy is
\begin{eqnarray}
  \frac{d F_{ch}}{dt} &=&  \int_V \frac{\partial f_{ch}(c)}{\partial c}\frac{\partial c}{\partial t} d^3x =  \int_V - \frac{\partial f_{ch}(c)}{\partial c} \nabla \cdot \vec{J}_c d^3x  \nonumber \\
                      &=&  - \oint_A \frac{\partial f_{ch}}{\partial c} \vec{J}_{c} \cdot \vec{n} da - \int_V  (-\nabla \mu_{c}) \cdot M_0 c(1-c) (-\nabla \mu_{c}) dv \nonumber \\
                      & & -  \int_V  (-\nabla \mu_{c}) \cdot M_0 c(1-c) (eZ_e\vec{E}) dv
\end{eqnarray}
where the first term is the chemical free energy conducted by the flux $\vec{J}_{c}$; the second term is the dissipation of the chemical free energy usually in the form of heat; the third term is the inner product between the chemical driving force and the coupling portion of the mass flux due to the effective driving force of electromigration, which represents the conversion between the electrical energy and the chemical energy via the electrical work. The polarity effect of electromigration is well known. The effective charge number is usually negative, thus the effective driving force is usually in the same direction with the electron flux. The mass flux lies in the same direction with the chemical driving force $-\nabla \mu_c$. Then when the electron flux lies in the opposite direction with the mass flux, then the third term is positive. This indicate the consumption of the chemical free energy is lowered. Since $\vec{E}=\rho \vec{J}$ and $Z_e\propto J$, then the effective driving force itself is proportional to the rate of the work $J^2\rho$ done by the electrical field. Then, in this case the electrical energy is converted into the chemical free energy via the electrical work, as shown by the third term so that the consumption of the chemical free energy is lowered. In the other case when the electron flux lies in the same direction with the mass flux, then the third term is negative. Thus, the consumption of the chemical free energy is enhanced and it is actually converted into the electrical energy.

In this subsection, the system is assumed to be close to equilibrium so that irreversible processes behave linearly. Analyses found that when the coupling dependence of energies on other independent variables is negligible, the law of energy conservation during the process of energy conversion leads to Onsager's relations directly. In the following subsection, nonlinear irreversible processes will be discussed using PFVA with the energy functionals. Using this approach, the fluxes, without resorting to any linear assumptions, can be well defined by guaranteeing the total free energy decreases in an isolated system as time evolves.

\subsection{Nonlinear Irreversible Processes using PFVA}

Consider the same system studied above. Here the free energy density functions are written as $\epsilon'_\psi(\psi, \eta)$ and $\epsilon'_\psi(\eta, \psi)$. The coupling dependence of energies on independent variables is explicitly shown and denoted by a prime. That is, here energies are assumed to be dependent on all independent variables. On contrast, in Onsager's approach, fluxes are assumed to be dependent on all driving forces of simple irreversible processes. This is an important difference between PFVA and Onsager's approach. However, the assumption of local thermodynamic equilibrium still needs to be used so that the energy density functions can be well defined within each volume element. Then in the system considered here, the rate of change of the total free energy is
\begin{eqnarray}
  \frac{d (E'_\psi+E'_\eta)}{dt} &=&  \int_V \{ [\frac{\partial (\epsilon'_\psi+\epsilon'_\eta)}{\partial \psi }] \psi_t + [\frac{\partial (\epsilon'_\psi+\epsilon'_\eta)}{\partial \eta }] \eta_t  \} d^3x \nonumber \\
   &=&  \int_V \{ \Lambda'' \psi_t + \Omega'' \eta_t  \} d^3x  =  \int_V \{ - \Lambda'' \nabla \cdot \vec{J}_{\psi} - \Omega'' \nabla \cdot \vec{J}_{\eta}  \} d^3x \nonumber \\
   &=& - \oint_A \{\Lambda'' \vec{J}_{\psi} + \Omega'' \vec{J}_{\eta} \}  \cdot \vec{n} da + \int_V \{ \nabla \Lambda'' \cdot \vec{J}_{\psi} + \nabla \Omega'' \cdot \vec{J}_{\eta}  \} d^3x,
\end{eqnarray}
where $\Lambda''$ and $\Omega''$ are introduced for brevity. To guarantee the total free energy decreases as time evolves in an isolated system, the fluxes can be defined as
\begin{eqnarray}
   \vec{J}_{\psi} &=& - M_\psi \nabla \Lambda'' = - M_\psi (\nabla \frac{\partial \epsilon'_\psi}{\partial \psi } + \nabla \frac{\partial \epsilon'_\eta}{\partial \psi } ) \label{eq:33}\\
     &=& - M_\psi \frac{\partial^2 (\epsilon'_\psi+\epsilon'_\eta)}{\partial^2 \psi } \nabla \psi - M_\psi \frac{\partial^2 (\epsilon'_\psi+\epsilon'_\eta)}{\partial \psi \partial \eta } \nabla \eta, \\
   \vec{J}_{\eta} &=& - M_\eta \nabla \Omega''  = - M_\eta (\nabla \frac{\partial \epsilon'_\eta}{\partial \eta } + \nabla \frac{\partial \epsilon'_\psi}{\partial \eta } ) \label{eq:34}\\
     &=& - M_\eta \frac{\partial^2 (\epsilon'_\psi+\epsilon'_\eta)}{\partial^2 \eta } \nabla \eta - M_\eta \frac{\partial^2 (\epsilon'_\psi+\epsilon'_\eta)}{\partial \eta \partial \psi } \nabla \psi,
\end{eqnarray}
where expansions are performed to enable a comparison with the fluxes defined in Onsager's approach. Eqns (\ref{eq:31}) and (\ref{eq:32}) can be rewritten as
\begin{eqnarray}
\vec{J}_{\psi} &=& -L_{\psi\psi} \frac{\partial^2 \epsilon_\psi}{\partial^2 \psi } \nabla \psi -L_{\psi\eta} \frac{\partial^2 \epsilon_\eta}{\partial^2 \eta} \nabla \eta,  \\
\vec{J}_{\eta} &=& -L_{\eta\eta} \frac{\partial^2 \epsilon_\eta}{\partial^2 \eta } \nabla \eta -L_{\eta\psi} \frac{\partial^2 \epsilon_\psi}{\partial^2 \psi} \nabla \psi.
\end{eqnarray}
Match of coefficients leads to
\begin{eqnarray}
 L_{\psi\psi} &=& M_\psi \frac{\partial^2 (\epsilon'_\psi+\epsilon'_\eta)}{\partial^2 \psi } / \frac{\partial^2 \epsilon_\psi}{\partial^2 \psi } \\
 L_{\psi\eta} &=& M_\psi \frac{\partial^2 (\epsilon'_\psi+\epsilon'_\eta)}{\partial \psi \partial \eta } / \frac{\partial^2 \epsilon_\eta}{\partial^2 \eta}  \\
 L_{\eta\eta} &=& M_\eta \frac{\partial^2 (\epsilon'_\psi+\epsilon'_\eta)}{\partial^2 \eta } / \frac{\partial^2 \epsilon_\eta}{\partial^2 \eta } \\
 L_{\eta\psi} &=& M_\eta \frac{\partial^2 (\epsilon'_\psi+\epsilon'_\eta)}{\partial \eta \partial \psi } / \frac{\partial^2 \epsilon_\psi}{\partial^2 \psi}.
\end{eqnarray}
Here it is reasonable to believe that the coupling dependence of energies is usually weak. Thus, $\epsilon'_\psi \approx \epsilon_\psi$ and $\frac{\partial^2 \epsilon'_\eta}{\partial^2 \psi }\ll \frac{\partial^2 \epsilon'_\psi}{\partial^2 \psi }$, then $L_{\psi\psi} \approx M_\psi$. Furthermore, $\frac{\partial^2 (\epsilon'_\psi+\epsilon'_\eta)}{\partial \psi \partial \eta }$ is usually much smaller than $\frac{\partial^2 \epsilon_\eta}{\partial^2 \eta}$, then $L_{\psi\eta}$ is usually one order of magnitude lower than $L_{\psi\psi}$. This is very plausible since the coupling effects are usually minor. Analogously, the same conclusions can be drawn on $L_{\eta\eta}$ and $L_{\eta \psi}$. Note that, using PFVA only one mobility coefficient is present in the expression of each flux, such as $M_\psi$ in $\vec{J}_\psi$ and $M_\eta$ in $\vec{J}_\eta$. These coefficients can be determined using material coefficients such as diffusivity and conductivity. Then the coupling terms in fluxes are fully determined. In Onsager's approach, there are two mobility coefficients in the expression of each flux. It was shown in the subsection above, Onsager's relations on the coupling coefficients guarantees the energy is conserved during the conversion of the two energies. Then, will the definition of fluxes using PFVA achieve this purpose? To address this problem, dissipation and conversion of energies are further analysed.

First, consider the rate of change of the energy $E'_{\psi}(\psi, \eta)$, and we have
\begin{eqnarray}
\frac{dE'_{\psi}}{dt} &=& \int_V [ \frac{\partial \epsilon'_{\psi}}{\partial \psi} \frac{\partial \psi}{\partial t} + \frac{\partial \epsilon'_{\psi}}{\partial \eta} \frac{\partial \eta}{\partial t} ] dv  \nonumber \\
 &=& - \int_V (\frac{\partial \epsilon'_{\psi}}{\partial \psi} \nabla \cdot \vec{J}_{\psi}+ \frac{\partial \epsilon'_{\psi}}{\partial \eta} \nabla \cdot \vec{J}_{\eta}) dv \nonumber \\
 &=& - \oint_A [ \frac{\partial \epsilon'_{\psi}}{\partial \psi} \vec{J}_{\psi} + \frac{\partial \epsilon'_{\psi}}{\partial \eta} \vec{J}_{\eta} ] \cdot \vec{n} da + \int_V  \nabla \frac{\partial \epsilon'_{\psi}}{\partial \psi} \cdot \vec{J}_{\psi} dv +  \int_V \nabla \frac{\partial \epsilon'_{\psi}}{\partial \eta} \cdot \vec{J}_{\eta} dv. \label{eq:35}
\end{eqnarray}
Substitution of the fluxes in Eqns (\ref{eq:33}) and (\ref{eq:34}) into the last two volume integrals above, after simplification we have
\begin{eqnarray}
\int_V  \nabla \frac{\partial \epsilon'_{\psi}}{\partial \psi} \cdot \vec{J}_{\psi} dv &+&  \int_V \nabla \frac{\partial \epsilon'_{\psi}}{\partial \eta} \cdot \vec{J}_{\eta} dv = \nonumber \\
 &-& \int_V \nabla \frac{\partial \epsilon'_{\psi}}{\partial \psi} \cdot M_\psi \nabla \frac{\partial \epsilon'_\psi}{\partial \psi } dv - \int_V \nabla \frac{\partial \epsilon'_{\psi}}{\partial \eta} \cdot M_\eta \nabla \frac{\partial \epsilon'_\psi}{\partial \eta } dv  \nonumber \\
 &-& \int_V  \nabla \frac{\partial \epsilon'_{\psi}}{\partial \psi} \cdot M_\psi \nabla \frac{\partial \epsilon'_\eta}{\partial \psi } dv -  \int_V \nabla \frac{\partial \epsilon'_{\psi}}{\partial \eta} \cdot M_\eta \nabla \frac{\partial \epsilon'_\eta}{\partial \eta } dv. \label{eq:36}
\end{eqnarray}
The first two terms represent the dissipation of energy $E'_{\psi}$ since the integrands include the inner product between the portion of driving force $-\nabla \frac{\partial \epsilon'_{\psi}}{\partial \psi}$ ($- \nabla \frac{\partial \epsilon'_{\psi}}{\partial \eta}$) and its conjugate portion of flux $- M_\psi \nabla \frac{\partial \epsilon'_\psi}{\partial \psi }$ ($- M_\eta \nabla \frac{\partial \epsilon'_\psi}{\partial \eta }$). The former is a major contribution to the dissipation of energy arising from the major portion of the flux $\vec{J}_{\psi}$. While the latter is a minor contribution arising from the coupling portion of the flux $\vec{J}_{\eta}$, which leads to the variation of the field $\eta$ and thus contributes to the dissipation of the energy $E_{\psi}$ via the coupling dependence $\frac{\partial \epsilon'_{\psi}}{\partial \eta}$. The last two terms represent the conversion between energies $E'_{\psi}$ and $E'_{\eta}$. The integrand of the third term is $\nabla \frac{\partial \epsilon'_{\psi}}{\partial \psi} \cdot M_\psi \nabla \frac{\partial \epsilon'_\eta}{\partial \psi }$. Here $-M_\psi \nabla \frac{\partial \epsilon'_\eta}{\partial \psi }$ is the coupling portion of the flux $\vec{J}_{\psi}$ and its driving force is $- \nabla \frac{\partial \epsilon'_\eta}{\partial \psi }$ arising from the energy density function $\epsilon'_\eta$. This coupling portion of the flux $\vec{J}_{\psi}$ contributed to the variation of the field $\psi$ and thus causes the variation of $\epsilon'_\psi$. In brief, in the third term, the driving force arising from $\epsilon'_\eta$ contributed to the variation of $\epsilon'_\psi$. Hence, this term represents the conversion between the energies $E'_{\eta}$ and $E'_{\psi}$. Analogously, in the fourth term, the driving force arising from $\epsilon'_\eta$ results in the major portion, $-M_\eta \nabla \frac{\partial \epsilon'_\eta}{\partial \eta }$, of the flux $\vec{J}_{\eta}$. This contributes to the variation of the field $\eta$ and thus the variation of the energy density function $\epsilon'_\psi$ via the partial derivative $\frac{\partial \epsilon'_{\psi}}{\partial \eta}$. As a result, the fourth term also represents the conversion between the energies $E'_{\eta}$ and $E'_{\psi}$.

The rate of change of the energy $E'_{\eta}(\eta, \psi)$ can be considered in the same way. Its expression is the same as those in Eqns (\ref{eq:35}) and (\ref{eq:36}) after the switch of the two subscript $\psi$ and $\eta$. And it is
\begin{eqnarray}
\frac{dE'_{\eta}}{dt'} &=& \frac{dE_{\eta}^{'cd}}{dt'} + \frac{dE_{\eta}^{'ds}}{dt'} + \frac{dE_{\eta}^{'cv}}{dt'},  \\
    \frac{dE_{\eta}^{'cd}}{dt'} &=& - \oint_A [ \frac{\partial \epsilon'_{\eta}}{\partial \eta} \vec{J}_{\eta} + \frac{\partial \epsilon'_{\eta}}{\partial \psi} \vec{J}_{\psi} ] \cdot \vec{n} da,   \\
    \frac{dE_{\eta}^{'ds}}{dt'} &=& - \frac{dQ}{dt} = - \int_V \nabla \frac{\partial \epsilon'_{\eta}}{\partial \eta} \cdot M_\eta \nabla \frac{\partial \epsilon'_\eta}{\partial \eta } dv - \int_V \nabla \frac{\partial \epsilon'_{\eta}}{\partial \psi} \cdot M_\psi \nabla \frac{\partial \epsilon'_\eta}{\partial \psi } dv,   \\
    \frac{dE_{\eta}^{'cv}}{dt'} &=& - \int_V  \nabla \frac{\partial \epsilon'_{\eta}}{\partial \eta} \cdot M_\eta \nabla \frac{\partial \epsilon'_\psi}{\partial \eta } dv - \int_V \nabla \frac{\partial \epsilon'_{\eta}}{\partial \psi} \cdot M_\psi \nabla \frac{\partial \epsilon'_\psi}{\partial \psi } dv. \label{eq:37}
\end{eqnarray}
Careful examination of the two terms on the right hand side (r.h.s.) of Eqn (\ref{eq:37}) and the last two terms on the r.h.s. of Eqn (\ref{eq:36}) indicates they are identical. That is, using a time reversal $t'=-t$ and following the same arguments presented in the above subsection, it is
straightforward to show that
\begin{equation}
\frac{dE^{'cv}_{\psi}}{dt} + \frac{dE^{'cv}_{\eta}}{dt} \equiv 0.
\end{equation}
Consequently, the definition of the two fluxes in Eqns (\ref{eq:33}) and (\ref{eq:34}) using PFVA with the energy functionals guarantees that the energy is conserved during the process of energy conversion. That is, the first law of thermodynamics is satisfied. Note that the odd and even variables with respect to time reversal is no considered, since odd variables usually appear in pairs in the energy functionals.

The Thomson effect is also a concrete example of conversion of energies because it indicates that a portion of electrical energy is converted into heat under the influence of an externally applied electrical field. In Eqn (\ref{eq:35}), let $\epsilon'_{\psi}=\int_0^T C_P(\theta, n) d \theta$ be the internal energy so that $\psi=T$ is the temperature and $\eta=n$ is the density of electrons. Then Eqn (\ref{eq:35}) becomes
\begin{eqnarray}
\frac{dE'_{T}}{dt} &=& - \oint_A [ \frac{\partial \epsilon'_{T}}{\partial T} \vec{J}_{q} + \frac{\partial \epsilon'_{T}}{\partial n} \vec{J}_{n} ] \cdot \vec{n} da + \int_V  \nabla \frac{\partial \epsilon'_{T}}{\partial T} \cdot \vec{J}_{q} dv +  \int_V \nabla \frac{\partial \epsilon'_{T}}{\partial n} \cdot \vec{J}_{n} dv,
\end{eqnarray}
where the third term on the r.h.s. can be shown to be related to the Thomson effect $q_t =-\kappa \vec{J}\cdot \nabla T$. Analyses is performed in metals so that $n$ is independent of $T$. The integrand in the third term can be expanded as
\begin{eqnarray}
 \nabla \frac{\partial \epsilon'_{T}}{\partial n} \cdot \vec{J}_{n} &=& \nabla \frac{\partial \int_0^T C_P(\theta, n) d \theta }{\partial n} \cdot \vec{J}_{n} \nonumber \\
  &=& [\frac{\partial^2 C_P(T,n)}{\partial T \partial n} \nabla T + \int_0^T \frac{\partial^2 C_P(\theta, n)}{\partial^2 n} d \theta \nabla n] \cdot (- \vec{J}) \\
  &\approx& - \frac{\partial^2 C_P(T,n)}{\partial T \partial n} \nabla T \cdot \vec{J},
\end{eqnarray}
where at the last step, $\frac{\partial^2 C_P(\theta, n)}{\partial^2 n}$ is assumed negligible and $\nabla n$ as well due to the high mobility of electrons. Then, the Thomson coefficient is found to be $\kappa=\frac{\partial^2 C_P(T,n)}{\partial T \partial n}$. With Eqns (\ref{eq:10}) and (\ref{eq:5}), the Thomson effect can be rewritten as
\begin{equation}
q_t = - \kappa \sigma S_e \nabla T \cdot \nabla T  - \kappa \sigma \vec{E} \cdot \nabla T  - \kappa \sigma  [\frac{R}{T \partial R/\partial T} (\frac{3}{2}+\frac{E_c-\mu}{k_B T})] \vec{E} \cdot \nabla T.
\end{equation}
The physical content of the above equation is straightforward. The first term is due to the coupling portion of electron flux, $-\sigma S_e \nabla T$, which arises from the thermal driving force $- \nabla T$. Since $S_e$ consists of the derivative of the heat capacity w.r.t the number density of electrons and this term involves $\nabla T \cdot \nabla T$, then this term represents the dissipation of heat. That is, when electrons are driven down the temperature gradient, heat is released by them to the surroundings, due to the difference of the heat carried by them at the two ends with higher and lower temperatures. Both the second and the third terms represent the energy conversions between the electrical energy and thermal energy. In the case when $\vec{E}$ and $\nabla T$ lie in opposite directions, the second term is positive and thus the rate of consumption of the thermal energy is lowered. The electron flux lies in the direction of $\vec{E}$, then electrons are driven up the temperature gradient in this case. Their electrical potential energy lowers while their thermal energy increases. Thus, the electrical potential energy of electrons is converted into the thermal energy of electrons, which leads to a lower consumption rate of the thermal energy of the system. As a result, to main a steady state of conduction of heat, heat is released from the system into the surroundings. The third term only exists in semiconductors. When electrons are driven up the temperature gradient, their number density in the conduction band is also increased. Then, a portion of the electric potential energy is also used to activate more electrons into the conduction band, and thus the electrical energy is converted into the thermal energy of the system by this way. In the other case when $\vec{E}$ and $\nabla T$ lie in the same direction, the second term is negative and thus the rate of consumption of the thermal energy is enhanced. In this case, electrons are driven down the temperature gradient. The temperature gradient provides an additional driving force for the conduction of electrons, thus the electrical work is saved. Thus, in this sense, the thermal energy can be said to be converted into the electrical potential energy of the system. As a result, to main a steady state of conduction of heat, heat needs be absorbed into the system from the surroundings.

In this section, the dissipation and conversion of energy are mainly discussed. Dissipation of energies usually lead to the increase of entropy since energies are usually dissipated as heat. However, the conversion of energies also leads to the increase of entropy. Consider the same thermodynamic
system above with $\textbf{S}_{\psi}$ and $\textbf{S}_{\eta}$ being the total entropies. Using PFVA with the entropy functionals, it is straightforward to
show that terms consisting of $\nabla \frac{\partial S_{\psi}}{\partial \psi} \cdot \nabla \frac{\partial S_{\eta}}{\partial \psi}$ and $\nabla \frac{\partial S_{\eta}}{\partial \eta} \cdot \nabla \frac{\partial S_{\psi}}{\partial \eta}$, which represent the conversion of energies, also contributes to the production of entropy. This is consistent with the definition of the rate of production of entropy according to Onsager's formulations, i.e., $\frac{\partial S}{\partial t} = \frac{1}{T} \displaystyle{\sum_i} J_i X_i = \frac{1}{T} \displaystyle{\sum_{ij}} L_{ij} X_i X_j$, where the coupling products are also present. Consequently, both dissipation and conversion of energies lead to the production of entropies.

For irreversible processes, both the energy and entropy functionals can be used with PFVA to construct governing equations. If both functionals were clearly defined and their relations were thoroughly known, then these two approaches should be equivalent. However, our understanding of entropy itself and the entropy functions for most irreversible processes are still quite insufficient. Thus, for isothermal processes,
the energy functionals are usually favorable owing to the explicit definition of the energy density functions for these processes. Furthermore, it is more convenient to relate the fluxes to the dissipation and conversion of energies. However, for processes coupled with thermal conduction, the entropy functionals are probably the only choice. Because, first, using the entropy functionals the second law of thermodynamics is strictly satisfied; second and more importantly, for thermal conduction process, it is quite bizarre there is no such definition as a free thermal energy function. Analyses for TMDT using the free energy functional with $\int_0^T C_P(\theta) d \theta - T \int_0^T \frac{C_P(\theta)}{\theta} d \theta$ representing the thermally related contribution leads to no clear results and neat interpretations. Though the concept of free energy triumphed over entropy as a more popular state function in thermodynamics, it is argued here that the role of entropy should become more important in the future. First, natural processes are never ideally isothermal. Energy is always dissipated and entropy generated. For examples, heat is released or absorbed during chemical reactions and so is latent heat during phase transformations; heat is also dissipated during both the electrical conduction and chemical diffusion. Variation of temperatures is very common during these processes. Thus, a thorough understanding of the dissipation and conversion of energy during these processes requires the use of the function entropy. Second, order-from-disorder is an important step during abiogenesis and other self-assembly processes, and entropy itself is directly related to orderliness. Thus, using entropy to analyse these processes is a more reasonable choice. It is a pity currently only the entropy functions of thermal conduction and chemical diffusion are well defined. Each process should be related to a entropy function as well as a free energy function. It is believed that the determination of these entropy functions can greatly benefit our understanding of irreversible processes.

\section{Diffusion under the Influence of Elastic Fields}

Diffusion under the influence of elastic fields is another important irreversible process which has been heavily investigated.$^{\cite{Chen:2002}}$ In a solid material, elastic strains usually arise either from the compositional inhomogeneity or the lattice mismatch between phases with different crystal structures. As a result, the diffusion process is naturally accompanied by the evolution of elastic fields and so is the variation of the chemical free energy by that of the elastic energy. However, the former is an irreversible process which abides by the law of thermodynamics; while the latter is an reversible process which abides by the law of classic mechanics. In the following discussions, PFVA is modified to incorporate a reversible process and each process abides by its corresponding law, respectively. Here, consider a binary system consists of two species with $c$ denotes the composition of one species. For simplicity, only elastic strains due to the compositional inhomogeneity is considered. The elastic strains here are assumed to be infinitesimal strains, thus the difference between the material derivative and the spatial derivative can be ignored.  Then, in this system, the coupling processes cause the variation of three energies: the kinetic energy of the elastic fields, $E_k=\int_V \epsilon_{k} dv= \int_V \frac{1}{2} \rho \nu_i \nu_i dv$ where $\nu_{i}$ is the velocity; the elastic energy, $E_{el}=\int_V \epsilon_{el} dv$; and the chemical free energy of diffusion, $E_{ch}=\int_V \epsilon_{ch} dv$. Thus, the rate of change of the total energy is
\begin{eqnarray}
\frac{d}{dt}(E_k+E_{ch}+E_{el}) &=& \frac{d}{dt} \int_V (\epsilon_{k}+\epsilon_{ch} + \epsilon_{el}) dv \nonumber \\
       &=& \int_V [\frac{\partial}{\partial t}(\frac{1}{2} \rho \nu_i \nu_i)+ (\frac{\partial \epsilon_{ch}}{\partial c} + \frac{\partial \epsilon_{el}}{\partial c}) \frac{\partial c}{ \partial t} + ( \frac{\partial \epsilon_{ch}}{\partial e_{ij}} + \frac{\partial \epsilon_{el}}{\partial e_{ij}} ) \frac{\partial e_{ij}}{ \partial t} ] dv \nonumber \\
       &=& \int_V [\rho \nu_i \frac{\partial \nu_i}{\partial t} - (\frac{\partial \epsilon_{ch}}{\partial c} + \frac{\partial \epsilon_{el}}{\partial c}) \nabla \cdot \vec{J}_c + ( \frac{\partial \epsilon_{ch}}{\partial e_{ij}} + \sigma_{ij} ) \nu_{i,j} ] dv \nonumber \\
       &=&- \oint_A ( \frac{\partial \epsilon_{ch}}{\partial c} + \frac{\partial \epsilon_{el}}{\partial c} ) \vec{J}_c \cdot \vec{n} da + \int_V \nabla (\frac{\partial \epsilon_{ch}}{\partial c} + \frac{\partial \epsilon_{el}}{\partial c}) \cdot \vec{J}_{c} dv \nonumber \\
       & & + \oint_A (\frac{\partial \epsilon_{ch}}{\partial e_{ij}} + \sigma_{ij} ) \nu_i n_j da + \int_V [\rho \frac{\partial \nu_i}{\partial t} - ( \frac{\partial \epsilon_{ch}}{\partial e_{ij}} + \sigma_{ij} )_{,j} ] \nu_{i} dv \label{eq:38}
\end{eqnarray}
where at the third step $\frac{\partial e_{ij}}{ \partial t}=\frac{1}{2}(\nu_{i,j}+\nu_{j,i})$ and the symmetry of indices $i$ and $j$ are used. Note that diffusion is an irreversible process which abides by the law of thermodynamics. Thus, to guarantee as time evolves the total free energy decreases due to the diffusion in an isolated system, the flux can be defined as
\begin{equation}
\vec{J}_c = - M_c \nabla(\frac{\partial \epsilon_{ch}}{\partial c} + \frac{\partial \epsilon_{el}}{\partial c}).
\end{equation}
However, the evolution of elastic fields is a reversible process which abides by the law of classic mechanics. The corresponding mechanical movements, which are denoted by the velocity $\nu_i$, affect both the kinetic energy $E_k$ and the generalized potential energy of the system. Then, at each instant, the rate of change of the kinetic energy and that of the generalized potential energy owing to this mechanical movements should sum up to zero according to the law of conservation of the mechanical energy. Examination of the last volume integral of Eqn (\ref{eq:38}) found this term reflects the rate of the change of the total energy due to the mechanical movements. Thus, the integrand of this term should be zero, which leads to the modified equation of motion:
\begin{equation}
\rho \frac{\partial \nu_i}{\partial t}=\sigma_{ij,j}+ (\frac{\partial \epsilon_{ch}}{\partial e_{ij}})_{,j}.
\end{equation}
Note there is an extra contribution to the stress due to the coupling dependence of the chemical free energy on the elastic strains. Without any coupling effects, this equation becomes Cauchy's equation of motion in the absence of body forces.

Now, the dissipation and conversion of energies can be discussed. The rate of change of the chemical free energy is
\begin{eqnarray}
\frac{dE_{ch}}{dt} &=&  \int_V ( \frac{\partial \epsilon_{ch}}{\partial c} \frac{\partial c}{ \partial t} +  \frac{\partial \epsilon_{ch}}{\partial e_{ij}} \frac{\partial e_{ij}}{ \partial t} ) dv \nonumber \\
       &=& - \int_V \frac{\partial \epsilon_{ch}}{\partial c}  \nabla \cdot \vec{J}_c dv +  \int_V \frac{\partial \epsilon_{ch}}{\partial e_{ij}} \nu_{i,j} dv \nonumber \\
       &=& - \oint_A \frac{\partial \epsilon_{ch}}{\partial c}  \vec{J}_c \cdot \vec{n} da + \int_V \nabla \frac{\partial \epsilon_{ch}}{\partial c} \cdot \vec{J}_{c} dv \nonumber \\
       & & + \oint_A \frac{\partial \epsilon_{ch}}{\partial e_{ij}}  n_j \nu_i da - \int_V (\frac{\partial \epsilon_{ch}}{\partial e_{ij}})_{,j} \nu_{i} dv \nonumber \\
       &=& - \oint_A \frac{\partial \epsilon_{ch}}{\partial c}  \vec{J}_c \cdot \vec{n} da - \int_V \nabla \frac{\partial \epsilon_{ch}}{\partial c} \cdot M_c \nabla \frac{\partial \epsilon_{ch}}{\partial c} dv - \int_V \nabla \frac{\partial \epsilon_{ch}}{\partial c} \cdot M_c \nabla \frac{\partial \epsilon_{el}}{\partial c} dv \nonumber \\
       & & + \oint_A \frac{\partial \epsilon_{ch}}{\partial e_{ij}} n_j \nu_i da - \int_V (\frac{\partial \epsilon_{ch}}{\partial e_{ij}})_{,j} \nu_{i} dv. \label{eq:40}
\end{eqnarray}
where at the last step, the first term is the flow of the chemical free energy $E_{ch}$ due to the compositional flux at the surface; the second term
represents the dissipation of $E_{ch}$ into heat; the third term represent the conversion of energies between $E_{ch}$ and $E_{el}$ via the diffusion process; in the fourth term, $\frac{\partial \epsilon_{ch}}{\partial e_{ij}} n_j$ acts as an extra contribution of the surface contraction and the whole term represents the extra work done on the surface due to this contribution; in the last term, $(\frac{\partial \epsilon_{ch}}{\partial e_{ij}})_{,j}$ acts as a body force so that this term represents the work done by it and thus $E_{ch}$ is converted into $E_{el}$ via this term. Similarly, the rate of change of the elastic energy is
\begin{eqnarray}
\frac{dE_{el}}{dt} &=& \int_V  ( \frac{\partial \epsilon_{el}}{\partial c} \frac{\partial c}{ \partial t} + \frac{\partial \epsilon_{el}}{\partial e_{ij}} \frac{\partial e_{ij}}{ \partial t} ) dv \nonumber \\
       &=& - \int_V  \frac{\partial \epsilon_{el}}{\partial c} \nabla \cdot \vec{J}_c  dv + \int_V  \sigma_{ij} \nu_{i,j} dv \nonumber \\
       &=& - \oint_A \frac{\partial \epsilon_{el}}{\partial c} \vec{J}_c \cdot \vec{n} da + \int_V \nabla \frac{\partial \epsilon_{el}}{\partial c} \cdot \vec{J}_{c} dv \nonumber \\
       & & + \oint_A \sigma_{ij} n_j \nu_i da - \int_V \sigma_{ij,j} \nu_{i} dv \nonumber \\
       &=& - \oint_A \frac{\partial \epsilon_{el}}{\partial c} \vec{J}_c \cdot \vec{n} da- \int_V \nabla \frac{\partial \epsilon_{el}}{\partial c} \cdot M_c \nabla \frac{\partial \epsilon_{el}}{\partial c} dv - \int_V \nabla \frac{\partial \epsilon_{el}}{\partial c} \cdot M_c \nabla \frac{\partial \epsilon_{ch}}{\partial c} dv  \nonumber \\
       & & + \oint_A \sigma_{ij} n_j \nu_i da - \frac{d}{dt} \int_V \frac{1}{2} \rho \nu_{i} \nu_{i} dv + \int_V (\frac{\partial \epsilon_{ch}}{\partial e_{ij}})_{,j} \nu_{i} dv \label{eq:39}
\end{eqnarray}
where at the last step, $\sigma_{ij,j}= - \rho \frac{\partial \nu_i}{\partial t} + (\frac{\partial \epsilon_{ch}}{\partial e_{ij}})_{,j}$ is substituted. In Eqn (\ref{eq:39}), the first term represents of the flow of the elastic energy $E_{el}$ at the surface due to the diffusional flux; the second term represents the dissipation of $E_{el}$ into heat; the third term is the conversion of energies between $E_{ch}$ and $E_{el}$ (Note the first three terms represent the variation of $E_{el}$ due to the irreversible process diffusion); the fourth term represents the rate of work done by the surface traction $\sigma_{ij} n_j$ at the surface; the fifth term represents the rate of the conversion of $E_{el}$ into the kinetic energy $E_k$ of the system; the last term is the work done by the body force $(\frac{\partial \epsilon_{ch}}{\partial e_{ij}})_{,j}$ mentioned above (note that the last three term represent of the variation of $E_{el}$ due to the reversible evolution of elastic fields). When the evolution of elastic fields is not coupled with diffusion, then only the fourth term and the fifth term exist in the r.h.s. of Eqn (\ref{eq:39}) and the rate of change of the elastic energy is related to the work done by the surface traction and the rate of its conversion into the kinetic energy of the system. However, the coupling with diffusion makes this process much more complex and interesting. Not only the elastic energy is converted into the chemical free energy and dissipated into heat via the irreversible diffusion process, but also work is done by the extra body force $(\frac{\partial \epsilon_{ch}}{\partial e_{ij}})_{,j}$ to convert the chemical free energy into the elastic energy via mechanical movements. Note that, during the conversion of energies, the law of conservation of energies is also satisfied. First, because in both Eqns (\ref{eq:40}) and (\ref{eq:39}), this term $ - \frac{\partial \epsilon_{el}}{\partial c} \cdot M_c \nabla \frac{\partial \epsilon_{ch}}{\partial c}$ which represents the energy conversion between $E_{ch}$ and $E_{k}$ exists and the performance of a time reversal ensure that energy is conserved during the irreversible diffusion process. Second, the work done by the extra body force $(\frac{\partial \epsilon_{ch}}{\partial e_{ij}})_{,j}$ is negative in Eqn (\ref{eq:40}) and positive in Eqn (\ref{eq:39}), which also guarantee energy is conserved during the conversion between $E_{ch}$ and $E_{k}$ owing to the reversible evolution of the elastic fields. Here, the performance of a time reversal is not required for reversible processes. During the construction of thermodynamic equations, the total free energy is guaranteed to decreases as time evolves. In addition, each single free energy term also decreases as time evolves, e.g., as shown in Eqns (\ref{eq:40}) and (\ref{eq:39}). However, in a real system with complex irreversible processes present, usually one free energy decrease while the other increases. Consequently, the performance of a time reversal is necessary to ensure the satisfaction of conservation of energy during the conversion of energies. However, in reversible process such as mechanical movements discussed above, the law of conservation of mechanical energy already guarantees as time evolves the kinetic energy increases while the potential energy decreases or vice versa. Thus, the performance of a time reversal is not needed for reversible processes.

In this section, the diffusion process under the influence of elastic fields is discussed. PFVA is modified to incorporate the law of classic mechanics during the construction of governing equations. During the discussion of the dissipation and conversion of energies, it is found that the law of conservation of energy is satisfied via both the irreversible diffusion process and the reversible evolution of the elastic fields.

\section{Preliminary Discussions on Nonequilibrium Thermodynamics}

In the above analyses of irreversible processes using PFVA, it is shown that the assumption of linearization is no longer needed. However, the system still needs be assumed to be at local thermodynamic equilibrium. Thus, the entropy and energy densities within each volume element can be approximated by their expressions at equilibriums. As a result, the above discussions are only valid when the system is near equilibrium. However,
in nonequilibrium thermodynamics, systems which are far from equilibrium are usually discussed. In these systems the assumption of local thermodynamic equilibrium no longer holds. Thus, in the following discussions, PFVA is modified to describe the evolution of physical fields in nonequilibrium systems. For simplicity, only one simple irreversible process involving the evolution of a physical field $\phi$ is analysed here. Generalization of the following analyse to complex irreversible processes is believed to be straightforward.

To begin with, the entropy density function $S$ within each volume element needs be identified. When the volume element is far from equilibrium, then besides the variable $\phi$, there is also a gradient ($\nabla \phi$) and a flux ($\vec{J}_{\phi}$) in $\phi$ within each element. Here, $\phi$ describes the static configuration
of the element, $\nabla \phi$ describes the strength of a perturbation to this static configuration; while $\vec{J}_{\phi}$ describes the dynamic configuration
of the element. Usually, in simple cases, $\vec{J}_{\phi}$ can be related to $\nabla \phi$ by some kinetic coefficients. However, given a certain $\nabla \phi$, $\vec{J}_{\phi}$ can be larger or smaller. Thus, here not only $\phi$ and $\nabla \phi$, but also $\vec{J}_{\phi}$ are treated as independent variables. That is, the entropy function of each volume element is identified as $S(\phi, \nabla \phi, \vec{J}_{\phi})$. The adoption of $\nabla \phi$ as an independent variable is previously used in the energy functionals for analysing spinodal decompositions$^{\cite{Cahn:1961}}$ and latter generalized to the entropy functionals$^{\cite{Penrose:1990}}$. It is evident that the entropy density $S$ then becomes solely determined by the variable $\phi$ when the whole system achieved equilibrium. The entropy functional thus can be treated and expanded analogously$^{\cite{Cahn:1961}}$ as
\begin{eqnarray}
\textbf{S} &=& \int_V S(\phi, \nabla \phi, \vec{J}_{\phi}) dv  \nonumber \\
   &=& \int_V [ - \frac{1}{2} K_{J} |\vec{J}_{\phi}|^2 + S_0(\phi) - \frac{1}{2} K_{\phi} |\nabla \phi|^2 ] dv. \label{eq:41}
\end{eqnarray}
where $- \frac{1}{2} K_{J} |\vec{J}_{\phi}|^2$ can be tentatively defined as the kinetic contribution to the entropy, $S_k$. Then the rate of change of $\textbf{S}$ is
\begin{eqnarray}
\frac{d\textbf{S}}{dt} &=& \int_V [ - K_{J} \vec{J}_{\phi} \cdot \frac{\partial \vec{J}_{\phi}}{\partial t}  + \frac{\partial S_0(\phi)}{\partial \phi} \frac{\partial \phi}{\partial t} - K_{\phi} \nabla \phi \cdot \nabla \phi_t ] dv. \nonumber \\
                       &=& \int_V (- K_{J} \vec{J}_{\phi} \cdot \frac{\partial \vec{J}_{\phi}}{\partial t}) dv + \int_V  (\frac{\partial S_0(\phi)}{\partial \phi} + K_{\phi} \nabla^2 \phi ) \frac{\partial \phi}{\partial t} dv -  \oint_{A} K_{\phi} \phi_t \nabla \phi \cdot \vec{n} da \nonumber \\
                       &=& \int_V [- K_{J} \frac{\partial \vec{J}_{\phi}}{\partial t} + \nabla (\frac{\partial S_0(\phi)}{\partial \phi} + K_{\phi} \nabla^2 \phi ) ] \cdot \vec{J}_{\phi} dv \nonumber \\
                       &+& \oint_{A} [K_{\phi} \phi_t \nabla \phi  - (\frac{\partial S_0(\phi)}{\partial \phi} + K_{\phi} \nabla^2 \phi)\vec{J}_{\phi} ] \cdot \vec{n} da,
\end{eqnarray}
where at the second step, $\frac{\partial \phi}{\partial t}=-\nabla \vec{J}_{\phi}$ is used. Thus, to guarantee a positive rate of production of entropy,
the flux can be assumed to be
\begin{eqnarray}
\vec{J}_{\phi} &=& M_{\phi} [- K_{J} \frac{\partial \vec{J}_{\phi}}{\partial t} + \nabla (\frac{\partial S_0(\phi)}{\partial \phi} + K_{\phi} \nabla^2 \phi )] \label{eq:42}
\end{eqnarray}
Then, the governing equation is found to be
\begin{equation}
M_{\phi} K_{J} \frac{\partial^2 \phi}{\partial t^2} + \frac{\partial \phi}{\partial t} = - \nabla \cdot M_{\phi} [ \nabla (\frac{\partial S_0(\phi)}{\partial \phi} + K_{\phi} \nabla^2 \phi )].
\end{equation}
Let $G_0(\phi) = U_0 - T S_0(\phi)$, then the above equation can be directly converted into the governing equation of the modified phase field crystal model.
The hyperbolic telegrapher equation of heat conduction can also be recovered following the above derivation and using $\phi=T$, $K_{\phi}=0$ and $S_0(T)=\int_0^T \frac{C_p(\theta)}{\theta} d\theta$. Previous analyses were based on the Gibbs equation and usually the coefficients are set as
functions of fluxes.$^{\cite{Jou:2010}}$ It is straightforward to show that our approach developed here is in fact equivalent. However, with our approach, it is more convenient to discuss the physical content of each term in the entropy density function. In Eqn (\ref{eq:41}), the first term reflects the kinetic contribution to the entropy. Note that it is associated with a negative sign. That is, a system with a larger flux has a smaller entropy, which indicates the system is more ordered. In latter discussions, this term will be shown to be related to the ``kinetic" energy within the volume element. The remaining two terms can be considered as the contribution to the entropy from the potential energy within the volume element. The second term is the static contribution to the entropy from the equilibrium state. If $S_0(\phi)$ is taken as the mixing entropy, then this term is directly related to the Maxwell entropy which describes the microscopic configuration of atoms at equilibrium. The third term reflects the contribution to the entropy from a perturbation to the equilibrium configuration of atoms within each volume element, which is usually measured by the gradient of the physical field. Both of these two terms can be considered as contributions from the potential energies since they closely relate to the static configuration of atoms at the microscopic level. While the first term is a kinetic contribution since it is relate to the flux, i.e., the generalized velocity. Thus, in brief, the entropy within a volume element can have both kinetic and potential contributions.

In the following discussions, using the flux$\vec{J}_{\phi}$ defined in Eqn (\ref{eq:42}), the conversion of energies in this system can be discussed. First, the flux equation needs be rewritten as
\begin{eqnarray}
\vec{J}_{\phi} + M_{\phi} K_{J} \frac{\partial \vec{J}_{\phi}}{\partial t} =  - \frac{M_{\phi}}{T} [ \nabla (\frac{\partial G_0(\phi)}{\partial \phi} - K_{\phi} T \nabla^2 \phi )],   \label{eq:43}
\end{eqnarray}
using $G_0(\phi) = U_0 - T S_0(\phi)$. The generalized potential energy is $G(\phi) = G_0(\phi)+\frac{1}{2} K_{\phi} T |\nabla \phi|^2$. Then, the rate of change of the total potential energy $\textbf{G}$ is
\begin{eqnarray}
\frac{d\textbf{G}}{dt} &=& \int_V [ \frac{\partial G_0(\phi)}{\partial \phi} \frac{\partial \phi}{\partial t} + K_{\phi} T \nabla \phi \cdot \nabla \phi_t ] dv. \nonumber \\
                       &=& \int_V  \nabla (\frac{\partial G_0(\phi)}{\partial \phi} - K_{\phi} T \nabla^2 \phi )  \cdot \vec{J}_{\phi} dv \nonumber \\
                       &+& \oint_{A} [- K_{\phi} T \phi_t \nabla \phi  + (\frac{\partial G_0(\phi)}{\partial \phi} - K_{\phi} T \nabla^2 \phi)\vec{J}_{\phi} ] \cdot \vec{n} da. \label{eq:44}
\end{eqnarray}
The volume integral at the second step above can be rewritten as
\begin{eqnarray}
\int_V  \nabla (\frac{\partial G_0(\phi)}{\partial \phi} - K_{\phi} T \nabla^2 \phi )  \cdot \vec{J}_{\phi} dv &=& \int_V  - \frac{T}{M_{\phi}} (\vec{J}_{\phi} + M_{\phi} K_{J} \frac{\partial \vec{J}_{\phi}}{\partial t})  \cdot \vec{J}_{\phi} dv \nonumber \\
                       &=&  - \int_V  [\frac{T}{M_{\phi}} \vec{J}_{\phi} \cdot \vec{J}_{\phi} +  \frac{\partial}{\partial t} ( \frac{1}{2} T K_{J} \vec{J}_{\phi} \cdot \vec{J}_{\phi})] dv, \label{eq:45}
\end{eqnarray}
where at the second step, $\frac{T}{M_{\phi}} \vec{J}_{\phi} \cdot \vec{J}_{\phi}$ in the first term is the rate of the potential energy dissipated into heat, i.e., $\frac{\partial Q}{\partial t}$ as analysed previously; in the second term, $\frac{1}{2} T K_{J} \vec{J}_{\phi} \cdot \vec{J}_{\phi}$ within the time derivative can be identified as the ``kinetic" energy $E_k$ within each volume element due to the presence of the flux $\vec{J}_{\phi}$. It can be seen this ``kinetic" energy $E_k$ gives rise to the kinetic contribution $S_k$ in the entropy function in Eqn (\ref{eq:41}). Here, $S_k = - \frac{\partial E_k}{\partial T}$, which has the same relation as that between the potential contribution to entropy and the potential energy. Furthermore, due to the presence of $E_k$ in the above equation, the potential energy no longer monotonously decreases when the system is insulated. Combination of Eqns (\ref{eq:44}) and (\ref{eq:45}) leads to
\begin{equation}
\frac{d\textbf{G}}{dt} + \frac{d\textbf{Q}}{dt} + \frac{d\textbf{E}_\textbf{k}}{dt} = 0, \label{eq:46}
\end{equation}
in an insulated system. Here, comparisons can be made with movements in classic mechanics. In the absence of $\frac{d\textbf{Q}}{dt}$, then Eqn (\ref{eq:46})
reflects the conversion between the potential energy and the kinetic energy, which is similar to pure mechanic movements, and the governing equation can be directly shown
to be hyperbolic. Due to the oscillation in the field variable $\phi$, the potential energy will increase with time when the kinetic energy is converted into the potential energy. In the other case when $\frac{d\textbf{E}_\textbf{k}}{dt}$ is absent, Eqn (\ref{eq:46}) reflects the dissipation of the potential energy into heat due to the irreversible thermodynamic processes, and the governing equation can be directly shown to be parabolic. In an insulated system, due to the second law of thermodynamics, the rate of change of the potential energy is non-increasing; thus, the dissipation of the potential energy into heat is irreversible, i.e., heat can never be spontaneously converted into the potential energy. Note that heat arises from the chaotic kinetic movements of atoms and molecules at the microscopic level. Thus, heat is in fact a chaotic kinetic energy. On contrast, the general kinetic energy is an energy with orderliness. However, in non-equilibrium thermodynamics, the governing equation shown above is evidently a combination of both. Thus, in system far from equilibrium, it is possible that the potential energy increases with time if the hyperbolic movements dominates during a certain stage. Moreover, it is shown that both the kinetic and potential contributions to entropy equal the negative partial derivative of the corresponding energy w.r.t temperature. In pure mechanic movements, when both the kinetic and the potential energy have no temperature dependence, then $S=0$. Thus, during the conversion of energies in these processes, $\frac{dS}{dt}=0$ so that the energy is never dissipated. But in non-equilibrium thermodynamics, usually the energy, especially the potential energy, has a strong temperature dependence. Thus, during these thermodynamic processes, the entropy $S\neq0$ and conversion of energies will lead to change of $S$. Then, the second law of thermodynamics comes into play and the dissipation of energy arises. In a sense, the temperature-dependence of the kinetic and potential energies involved in a natural process can be used to a necessary condition to tell whether the process is reversible or not.

In this section, a preliminary discussion on non-equilibrium thermodynamics is provided. In brief, a kinetic contribution to the entropy density function $S$ is introduced and $S$ is written as $S(\phi, \nabla \phi, \vec{J}_{\phi})$. In a non-equilibrium system with multiple fields present, the total entropy density function can be written as $- \displaystyle{\sum_i} \frac{1}{2} K_{J} |\vec{J}_{\phi_i}|^2 + \displaystyle{\sum_i} S_{0i}(\phi_1,\phi_2,...) - \displaystyle{\sum_i} \frac{1}{2} K_{\phi_i} |\nabla \phi_i|^2$. Using PFVA with the entropy functional, governing equations can be constructed to abide by the second law of thermodynamics. With the energy functional, the dissipation and conversion of energies can be analysed and the satisfaction of the first law of thermodynamics can be shown following the analyses in previous sections. Thus, analyses using PFVA can also be generalized to study non-equilibrium thermodynamics.

\section{Summary}

Irreversible processes are very common in nature. A simple irreversible process is usually described by the evolution of an independent field variable. In a system with multiple fields present, these simple irreversible processes will interact and thus lead to complex irreversible processes. In this paper, PFVA with either the entropy or energy functional is used to analyse complex irreversible processes. For each simple irreversible process, it is associated with a entropy or free energy density function and each function contributes to the total entropy or energy functionals of the system. To account for the coupling effects which give rise to the complex irreversible processes, it is assumed that the entropy or free energy density function of each field depends on all independent field variables. That is, all entropy and energy density functions are field-interdependent. Then, using PFVA, the flux of each field can be determined by ensuring either the total entropy increases or the total free energy decreases as time evolves in an isolated system. Within the expressions of fluxes, contributions from the coupling effects can be identified and important kinetic coefficients can be obtained. PFVA with the entropy functionals is used to analyse examples such as TE effects and TDMT. Important kinetic coefficients found for the TE effects are
\begin{eqnarray}
 S_e &=& - \frac{R}{e T \partial R/\partial T} \frac{\partial C_P}{\partial n}, \nonumber \\
\Pi &=& - K T^2 ne (\frac{1}{C_P^2}\frac{\partial C_P}{\partial T} + \frac{1}{C_P}\frac{\partial^2 R}{\partial T^2}), \nonumber \\
\kappa &=& \frac{\partial^2 C_P(T,n)}{\partial T \partial n}; \nonumber
\end{eqnarray}
and those for TMDT are
\begin{eqnarray}
S_T &=& -\frac{1}{\rho_0 k_B T} \frac{\partial C_P}{\partial c} + \frac{1}{k_B} \frac{\partial^2 \Delta S^{ex}_{mix}}{\partial c \partial T }, \nonumber \\
Q^* &=& - T \frac{\partial C_P}{\partial c} + \rho_0 T^2 \frac{\partial^2 \Delta S^{ex}_{mix}}{\partial c \partial T }. \nonumber
\end{eqnarray}
Analyses on the driving forces for the TE effects show that these effects are related to the energy conversion between the thermal energy and the electrical energy, and likewise, TDMT is related to the conversion of heat into the chemical free energy of the system. In general, complex irreversible processes arise from the conversion between energies associated with each simple irreversible processes. To understand the dissipation of conversion of energies, PFVA with the energy functional is used to analyse the coupling effects in a general system with two physical fields present. The linear irreversible processes are analysed with the Onsager approach and the nonlinear irreversible processes with PFVA. In both approaches, the dissipation and the conversion of energies are explicitly determined and thus are available for numerical evaluation. It is also shown that both the Onsager's relations and the fluxes defined using PFVA guarantee the satisfaction of the first law of thermodynamics during the process of conversion of energies. Furthermore, diffusion under the influence of elastic fields is analysed and thus PFVA is modified to incorporate the evolution of elastic fields which is a reversible process. The diffusion process abides by the law of thermodynamics while the evolution of elastic fields abides by the law of classic mechanics. Analyses found during the process of conversion of energies, the law of conservation of energy is also satisfied via both the irreversible diffusion process and the reversible evolution of elastic fields. In the end, PFVA is generalized to obtain governing equations for a nonequilibrium thermodynamic system using an extra kinetic contribution to the entropy density function. The analyses can be extended to a nonequilibrium thermodynamic system with multiple physical fields present.

As a very popular approach, PFVA is initially developed to construct governing equations to guarantee that the second law of thermodynamics or its corollary is satisfied. Now, it is shown that the first law of thermodynamics is also satisfied using this approach. Thus, PFVA can be developed into a very practical tool to construct governing equations which satisfy both the first and the second law of thermodynamics. Though classic thermodynamics mainly studies the equilibrium of a thermodynamic system. However, in essence, the first and second laws are relations about the rate of change of energies and entropies, i.e., the dynamic features of a system. Consequently, the fully exploitation of PFVA has the potential of not only significantly advancing our understanding of the thermodynamics of irreversible processes, but also making thermodynamics as a discipline and the study of it truly dynamic.

\section{Acknowledgement}
We gratefully acknowledge the financial support from the National Natural Science Foundation of China (Grant No. 51201049).
PZ thanks the interdisciplinary EP program and WCJ at MSE@UVA, advises from Dr. F.-Q. Yang of CME@U. Kenturky and
Dr. D. Wei of MSE@THU, PRC, and helpful discussions about TE effects with Dr. L.-L. Li at MSE@THU.


\newpage

\begin{thebibliography}{99}
\bibitem{DeHoff:1993}
R. T. DeHoff, Thermodynamics in Materials Science, McGraw-Hill, (1993).
\bibitem{Prigogine:1967}
I. Prigogine, Thermodynamics of Irreversible Processes, Interscience, (1967).
\bibitem{Ziman:1999}
J. M. Ziman, Principles of the Theory of Solids, Cambridge University Press, (1999).

\bibitem{Onsager:1932}
L. Onsager, Phys. Rev., 37, 405, (1931).

\bibitem{Chen:2002}
L.-Q. Chen, Annu. Rev. Mater. Res., 32, 113, (2002).
\bibitem{Cahn:1961}
J.W. Cahn, Acta. Metall., 9, 795, (1961).
\bibitem{Penrose:1990}
O. Penrose and P. C. Fife, Physica D, 43, 44, (1990).
\bibitem{Wang:1993}
S.-L. Wang, R.F. Sekerka, A.A. Wheeler, B.T. Murray, S.R. Coriell, R.J. Braun and G.B. McFadden, Physica D, 69, 189, (1993).
\bibitem{Wheeler:1996}
A.A. Wheeler, G.B. McFadden and W.J. Boettinger, Proc. R. Soc. Lond. A, 452, 495, (1996).

\bibitem{Tritt:2011}
T. Tritt, Annu. Rev. Mater. Res., 41, 433-48, (2011).
\bibitem{DiSalvo:1999}
F. DiSalvo, Science, 285, 703, (1999).
\bibitem{Bell:2008}
L. Bell, Science, 321, 1457, (2008).
\bibitem{Mott:1979}
N. F. Mott and E. A. Davis, Electronic Processes in Non-Crystalline Materials, Oxford University Press, (1979).

\bibitem{Dommelen:2014}
L. van Dommelen, Quantum Mechanics for Engineers, http://www.eng.fsu.edu/$\sim$dommelen /quantum/.

\bibitem{Bulusu:2008}
A. Bulusu and D.G. Walker, Superlattices and Microstructures, 44, 36, (2008).
\bibitem{Sootsman:2009}
J. R. Sootsman, D. Y. Chung, and M. G. Kanatzidis, Angew. Chem. Int. Ed., 48, 8616, (2009).


\bibitem{Zheng:2002}
F. Zheng, Advances in Colloid and Interface Science, 97, 255, (2002).

\bibitem{Piazza:2008}
R Piazza and A Parola, J. Phys.: Condens. Matter, 20, 153102, (2008).

\bibitem{Wiegand:2004}
S. Wiegand, J. Phys.: Condens. Matter 16, 357, (2004).

\bibitem{Huntington:1975}
Huntington, H. B., Diffusion in Solids: Recent Developments,(ed. A. S. Nowick and J. J. Burton), (1975)


\bibitem{Tyndall:1870}
J. Tyndall, Proc. R. Inst., 6, 1, (1870).

\bibitem{Ludwig:1856}
C. Ludwig, Sitz. ber. Akad. Wiss. Wien Math.-Nat. wiss. Kl., 20, 539, (1856).
\bibitem{Soret:1879}
C. Soret, Arch. Sci. Phys. Nat., 2, 48, (1879).

\bibitem{Wurger:2006}
A. W$\ddot{u}$rger, Europhys. Lett., 74, 658, (2006).

\bibitem{Platten:2006}
J. K. Platten, Journal of Applied Mechanics, 73, 5, (2006).

\bibitem{Yang:2012}
M. Yang and M. Ripoll, J. Phys.: Condens. Matter, 24, 195101, (2012).

\bibitem{Buhr:2006}
S. Duhr and D. Braun, Proc. Natl. Acad. Sci. USA, 103, 19678 (2006).

\bibitem{Ye:2003}
H. Ye, C. Basaran and D. Hopkins, Appl. Phys. Lett., 82, 1045, (2003).

\bibitem{Hehenkamp:1976}
T. Hehenkamp, Electro- and Thermo-Transp. Met. Alloys, 68, (1976).

\bibitem{Philibert:1991}
J. Philibert, Atomic Movement: Diffusion and Mass Transport in Solids: Les Editions de Physique, (1991).

\bibitem{Talbot:1980}
L. Talbot, R.K. Cheng, R.W. Schefer, D.R. Willis, J. Fluid Mech. 101, 737, (1980).

\bibitem{Kittel:1986}
Kittel, Charles, Introduction to Solid State Physics, John Wiley \& Soms, Inc., (1986).

\bibitem{Brandes:1992}
E. A. Brandes and G. B. Brook, Smithells Metals Reference Book, Butterw orth-Heinemann, (1992).

\bibitem{Chuang:2006}
Y. C. Chuang and C. Y. Liu, Appl. Phys. Lett., 88, 174105 (2006).
\bibitem{Hsiao:2009}
H.-Y. Hsiao and C. Chen, Appl. Phys. Lett., 94, 092107, (2009).

\bibitem{Carlos:2005}
C. Nieto-Draghi, J. B. $\acute{A}$valos and B. Rousseau, J. Chem. Phys., 122, 1, (2005).

\bibitem{Derjaguin:1965}
B. V. Derjaguin, Y. V. Yalamov, J. Colloid Sci., 20, 555, (1965).

\bibitem{Brock:1962}
J. R. Brock, J. Colloid Sci., 17, 768, (1962).

\bibitem{Braun:2006}
S. Duhr and D. Braun, PNAS, 103, 52, (2006).

\bibitem{Miller:1960}
Miller DG, Chem Rev, 60:15¨C37, (1960).

\bibitem{Truesdell:1984}
Truesdell C, Rational Thermodynamics. 2nd edn. Springer, New York, (1984).

\bibitem{Jou:2010}
David Jou, Jos¨¦ Casas-V¨¢zquez and Georgy Lebon, Extended Irreversible Thermodynamics Fourth Edition, Springer, (2010).

\bibitem{Zhou:2011}
P. Zhou, W.C. Johnson, J. Electron. Mater., 40, 1867, (2011).



\end{thebibliography}
\end{document}